\documentclass[journal=jacsat,manuscript=article]{achemso}

\usepackage{graphicx}           
\usepackage[colorlinks,citecolor={blue}]{hyperref}

\usepackage{amsmath,amssymb}

\usepackage{bm,bbm}

\usepackage[version=3]{mhchem} 
\usepackage{xr}                 
\usepackage{xcolor}             
\usepackage[T1]{fontenc}        
\usepackage[normalem]{ulem}
\usepackage{multirow}

\newcommand{\msd}[1]{{\color[HTML]{0088ff} [[#1]]}}
\newcommand{\edits}[1]{#1}
\newcommand{\remove}[1]{}

\newcommand{\SI}[1]{#1}

\newcommand*{\strain}{u}
\newcommand*{\bstrain}{\mathbf{u}}

\author{Krista G.~Schoonover}
\affiliation{Department of Chemistry, Texas A\&M University, College Station, TX 77843, USA}
\author{Gaurav Rawat}
\affiliation{Department of Materials Science and Engineering, Texas A\&M University, College Station, TX 77843, USA}
\author{Emily B.~Pentzer}
\affiliation{Department of Chemistry, Texas A\&M University, College Station, TX 77843, USA}
\alsoaffiliation{Department of Materials Science and Engineering, Texas A\&M University, College Station, TX 77843, USA}
\author{Michael S.~Dimitriyev}
\affiliation{Department of Materials Science and Engineering, Texas A\&M University, College Station, TX 77843, USA}
\email{msdim@tamu.edu}

\title[Relaxation and elasticity of BCPs]{Structural Relaxation and Anisotropic Elasticity of Ordered Block Copolymer Melts}

\begin{document}

\begin{abstract}
Block copolymer (BCP) melts play a critical role in the design of thermoplastics, owing in large part to the creation of alternating nano-scale domains of soft and stiff components.
\edits{Considerable attention has been given to the short-to-intermediate time response of BCP melts, when the storage modulus is expected to dominate the viscoelastic properties.}
\remove{Much of thermoplastic design has been focused on the short-time response associated with the dynamical rigidity of rubbery or glassy chains.}
However, less attention has been paid to the long-time relaxation and rigidity of microphase separated BCP melts or the role that domain morphology plays in modulating near-equilibrium response.
We take advantage of the ability of self-consistent field theory (SCFT) to calculate equilibrium properties of BCP melts to explore the anisotropic elastic response of ordered ABA and AB copolymer melts as quasistatic deformation processes.
This allows us to determine the anisotropic stiffness of the liquid crystal-like lamellar and columnar phases due to modulations in domain spacing, as well as the full stiffness tensor of the cubic BCC sphere and double gyroid phases. 
We explore elastic modulus landscapes for both single grain materials and random polycrystals over architectural parameters and segregation strengths, using AB diblock and ABA triblock melts as key examples. 
\edits{Comparing AB and ABA melts, we show slight differences in equilibrium rigidity, which are shown to be due to a difference in effective domain segregation.}
\edits{Finally, we consider the equilibrium bending stiffnesses of lamellae and columnar phases, showing that the columnar phase is significantly stiffer than lamellae, with a characteristic bending length scale that is an order of magnitude larger.}
\end{abstract}

\section{Introduction}

Polymer melts play a central role in the landscape of polymer materials owing to their viscoelastic properties, which largely derive from chain entanglements.
Despite their ability to store stress over prolonged timescales, they are essentially polymer liquids and thus ultimately flow under applied stress, with their terminal relaxation determined by the time it takes for polymers to reptate past each other~\cite{Doi1986}.
The essential fluidity of polymer melts creates challenges for blended melts due to the propensity for unlike polymers to undergo phase separation; block copolymers (BCPs) would then play the role of compatibilizers, stabilizing blends such that the melt maintains microscopic interfaces creating complex structure that persists for long times~\cite{Jeong2025}.
Meanwhile, BCP thermoplastics take advantage of microphase separation between chemically-distinct blocks, resulting in stable, nano-patterned composites~\cite{Ruzette2005_Nature}.
Moreover, the nanostructure is self-assembled, with domain morphology arising from a careful balance of entropic costs due to chain stretching and enthalpic costs coming from block mixing at the intermaterial dividing surface (IMDS)---the interface between domains.
The equilibrium phases are periodic and capable of long range order, with predominant equilibrium phases of lamellae (1D layers), cylinders (2D, often hexagonally-packed), spheres (3D, often arranged on a BCC lattice), and bicontinuous networks, though exceptions exist due to blended additives or different polymer architectures~\cite{Matsen1999,Matsen2002}.
The bicontinuous network phases are of particular interest since they consist of labyrinthine domains and IMDSs that span all of the material volume while maintaining a unit cell with predominantly cubic symmetry.
Of the network phases, the double gyroid (DG) phase is the most common, and is thought to be optimal compared with competitor double diamond (DD) and double primitive (DP) phases for linear BCP architectures~\cite{Matsen1996,Dimitriyev2023}.

\begin{figure}[t]
    \centering
    \includegraphics[width=0.8\linewidth]{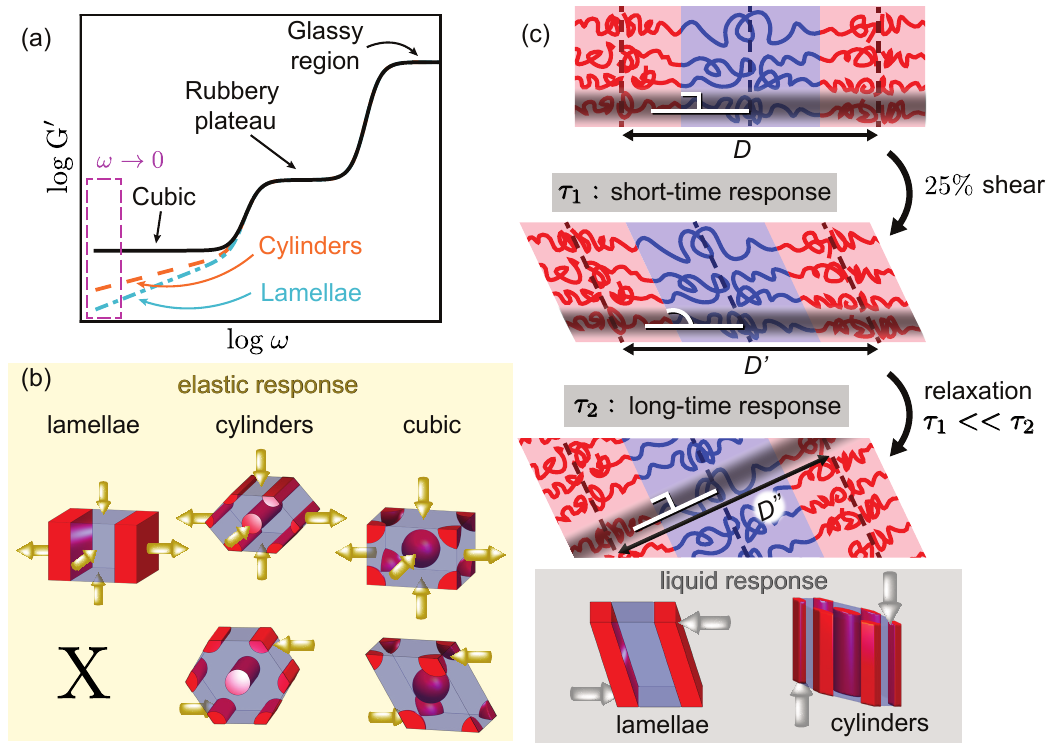}
    \caption{(a) Schematic of relative modulus of different phases for BCPs, this work narrowing in on $\omega \to 0$. Anisotropic elasticity is highlighted, showing deformations yielding liquid and elastic responses. (b) The elastic responses under investigation in this work (uniaxial extension and shear). (c) The short and long time response of a shear deformation is shown, highlighting the liquid response for lamellae and cylinders.}
    \label{fig:scheme_def_overview}
\end{figure}

The equilibrium phases formed by BCPs mirror those found in assemblies of surfactants as well as small-molecule liquid crystals and conventional crystals. 
Lamellar structures are stacks of parallel interfaces that resemble a smectic A phase, breaking translational symmetry in one dimension, whereas hexagonal packings of cylinders resemble a columnar phase, breaking translational symmetry in two dimensions.
Triply-periodic sphere and network phases break translational symmetry in all three dimensions and are thus structurally similar to crystalline solids.
Such ``crystalline liquids'' have unit cells on the scale of $10^2-10^3$ nm that contain thousands of polymer chains, making them much softer than their atomic and molecular crystal counterparts.
While the glass and rubber plateau moduli give the characteristic short-time responses due to slow polymer relaxation and entanglements and have been well-characterized, there are open questions regarding the mechanics and dynamics of BCP melts at terminal timescales\edits{, where the melt can be considered an equilibrium liquid}.
Kossuth, et al.~\cite{Kossuth1999} presented a schematic of how BCP melt stiffness changes as a function of timescale, as measured in an oscillatory rheometer (reproduced in Fig.~\ref{fig:scheme_def_overview}(a)).
While lamellar (1D) and cylinder (2D) phases are expected to flow over long times, with the cylinder phase characterized by a greater storage modulus ($G'$) than the lamellar phase, the 3D phases retain their rigidity for arbitrarily long time, as long as they are well-ordered.
This significant difference can be rationalized by the liquid crystalline structure of the 1D and 2D phases where the liquid behavior dictates the loss of rigidity at long timescales, whereas the 3D phases have crystalline order and thus their rigidity is a fundamental consequence of breaking continuous translational symmetry in all dimensions~\cite{Chaikin_Lubensky_1995}.
Nevertheless, the picture is complicated by the distribution of grain orientations in a ``polycrystalline'' melt and defect mechanics~\cite{Read1999,Knoll2002,Jinnai2006,Mareau2007,Kim2010,Pezzutti2011,Ryu2013,Li2015,Feng2021,Feng2023}, as well as a recently-reported linear instability of cubic DD under extensional deformation~\cite{Dimitriyev2025}.
Beyond such symmetry-based heuristics, little is known about the dependence of this \edits{equilibrium} rigidity on polymer composition and architecture, despite the large diversity of suitable systems and interests in thermoplastic design for desired applications.

In this Article, we examine how the \edits{equilibrium} rigidity of ordered BCP melts are afforded by conformational rigidity asymmetry \edits{$\epsilon \equiv \ell_{\rm A}/\ell_{\rm B}$ ($\ell_\alpha$ being the statistical segment length of block $\alpha)$}, polymer sequence (ABA vs.~AB), and the microphase separated pattern (see Fig.~\ref{fig:scheme_def_overview}).
By using SCFT to calculate equilibrium properties of BCP melts, we extract the \edits{equilibrium} moduli for lamellar, cylindrical, BCC sphere, and network (i.e., double gyroid) phases of symmetric ABA melts.
We show that the stiffness of these melts depart from intuition provided by standard composite theory, i.e.,~the rule of mixtures.
Instead, the extracted moduli show a non-monotonic dependence on composition, with effects arising from both nanoscale patterning and block sequence.
The block sequence dependence is rationalized though the domain structure of phase separated melts, wherein ABA molecules can possess both ``bridging'' conformations that join neighboring minority inclusions through the majority matrix, as well as ``looping'' conformations.
We calculate the polycrystalline average shear moduli, establishing upper (Voigt) and lower (Reuss) bounds for each phase~\cite{Voigt1889,Reuss1929,Hill1952}, providing a hierarchy of expected moduli and confirming that while 1D and 2D phases have vanishing lower bound moduli, the 3D phases have non-zero lower bounds.
By comparing with analogous AB diblock systems with comparable domain spacing, we demonstrate that while both architectures exhibit a similar range of \edits{equilibrium} moduli, there are small, yet significant differences.
\edits{We explore the differences between AB and ABA melt stiffness, finding that special care must be taken to distinguish between comparisons at fixed domain spacing and fixed segregation strength, highlighting the difference in an \emph{effective} degree of segregation due to chain architecture.}
\edits{Finally, we calculate quasistatic bending stiffnesses of lamellar and cylinder phases, showing that self-consistent field theory can be used to directly calculate healing lengths analogous to those of smectic A and columnar phases.}
\section{\edits{Equilibrium} rigidity of neat melts}


The ``\edits{equilibrium} rigidity'' of BCPs is inherently different from unstructured melts or blends.
For homopolymer melts, tensile stress elongates and aligns chains, creating internal stresses that are relieved over the disentanglement timescale~\cite{OConnor2019}.
Such melts retain essentially disordered arrangements after relaxing to an equilibrium state: the free energy per chain remains unchanged.
By combining multiple chemically-distinct monomers (A vs B), the equilibrium free energy acquires a free energy of mixing $F_{\rm mix}$ contribution.
It is variations to the free energy of mixing that can give rise to rigidity for arbitrarily long times.
For polymer blends, the equilibrium state consists of macrophase separated homopolymer domains, where each macroscopic domain has a locally vanishing free energy of mixing.
In this case, $F_{\rm mix} \approx \gamma A$, where $\gamma$ is an interfacial free energy density and $A$ is the interfacial area; in equilibrium, the interfacial area is minimized.
The only way for deformations to change the equilibrium mixing free energy of a blend is if the area of the interface between immiscible components increases even after relaxation.
Regardless if this can be achieved, the deformation free energy $\Delta F_{\rm mix}$ would still scale with interface area and thus be sub-extensive with system size, so that any measured stiffness would not be a bulk property of the melt.
By contrast, \emph{microphase separated} melts involve a mixing free energy that is extensive with system size: $F_{\rm mix} \sim V$, where $V$ is the melt volume.
If the equilibrium state of the melt corresponds to a stable minimum of the free energy of mixing, then any deformation from that equilibrium state will raise the value of $F_{\rm mix}$; within the linear response approximation, the change in free energy is given by
\begin{equation}
    \Delta F_{\rm mix} = \frac{1}{2}V\,C_{ijk\ell}\strain_{ij}\strain_{k\ell}\, ,
\end{equation}
where $C_{ijk\ell}$ is the linear elastic modulus tensor and $\strain_{ij}$ are components of the macroscopic strain tensor $\bstrain$.
Since the melt is taken to be incompressible, any change in volume must accompany a change in the number of polymer chains within that volume.
By considering deformations such that the trace of the strain tensor vanishes (${\rm tr}\,\bstrain = 0$), we only consider deformations that maintain a fixed melt volume, fixing the number of chains as expected for a realistic polymer melt.

The question of whether a phase is at a stable free energy minimum can be, to some degree, addressed through the symmetry of the phase in question, namely whether it has directions of continuous translational symmetry.
For the 1D lamellae, translational symmetry is broken in one dimension, leaving two symmetric directions along which we expect liquid-like behavior; for 2D cylinders, there is only one remaining translationally-symmetric direction.
The 1D and 2D phases are therefore stable with respect to deformations that change their $D$-spacing, but lack rigidity under certain deformations, resulting in certain zero-valued components of the linear elastic stiffness tensor $C_{ijk\ell}$.
The intuition behind this is exemplified in Fig.~\ref{fig:scheme_def_overview}(c).
As shown, shearing a lamellar structure will tilt chains relative to the IMDS for short times, incurring an entropic penalty.
Over long times, if allowed, the chains will flow, re-orienting the IMDSs such that the net effect of the deformation is a rotation of the melt.
Since it is possible for the initial $D$-spacing to be equal to its relaxed value $D''$, such a shear deformation can be performed without increasing the free energy per chain; thus the lamellar phase lacks a \edits{equilibrium} shear rigidity but has an extensional rigidity.

To compute the stiffness tensor components $C_{ijk\ell}$ for each phase, we take advantage of the ability of SCFT to rapidly calculate the equilibrium mixing free energy per chain.
It does this by solving the single-chain partition function within a mean field approximation where the chemical potential is self-consistently determined by single-chain conformations~\cite{Helfand1975,Matsen2002}.
Since this method immediately produces equilibrium predictions, it overcomes the time constraints on molecular dynamics approaches to simulating equilibrium states.
Moreover, advances in computing methods, particularly through the use of Anderson acceleration implemented in the open-source \texttt{PSCF} software~\cite{arora_broadly_2016}, have made rapid iteration over multiple phases, polymer architectures, and compositions feasible.
Thus, while prior work has either focused on lamellae~\cite{Thompson2004,Zhang2009,Zhu2011,Chu2024simulation}, cylinders~\cite{Chen2021}, or 3D phases~\cite{Kossuth1999,Tyler2003} for single polymer architectures, we opt to consider a more extensive set of structures and molecular parameters.
Recently, SCFT-imposed quasistatic deformations were used to study non-affine relaxation modes in bicontinuous networks~\cite{Dimitriyev2024} and demonstrate the instability of cubic DD under extensional strain~\cite{Dimitriyev2025}.

In our approach we first determine the equilibrium unit cell lattice parameters for a symmetric reference structure, then choose a strain protocol, update the lattice parameters, and finally iterate the SCFT equations until the residual error is minimized, obtaining a mixing free energy at fixed strain.
The elastic moduli are then determined by fitting a quadratic form to $\Delta F(\mathbf{\strain})$, giving $\Delta F$ in units of $n_{\rm ch}k_{\rm B}T$, where $n_{\rm ch} = \rho/N$ is the chain density with $1/\rho$ representing the volume occupied by a statistical segment and $N$ the degree of polymerization. 
This procedure is discussed in greater detail in \SI{Section S1}.
As established in prior studies~\cite{Matsen1999}, we represent symmetric ABA copolymers as ``doubled'' AB copolymers.
Consequently, the degree of polymerization $N^{\rm ABA}$ for ABA copolymers is taken to be twice that of AB diblocks, i.e.,~$N^{\rm ABA} = 2 N^{\rm AB}$.
If the segment density $\rho$ of both situations is identical, then the density of ABA chains is taken to be half the density of AB chains, i.e.,~$n^{\rm ABA}_{\rm ch} = n^{\rm AB}_{\rm ch}/2$.
This representation allows us to directly compare ABA triblocks to analogous systems of AB diblocks.


\subsection{1D: Lamellar rigidity}
\begin{figure}[t]
    \centering
    \includegraphics[width=0.8\linewidth]{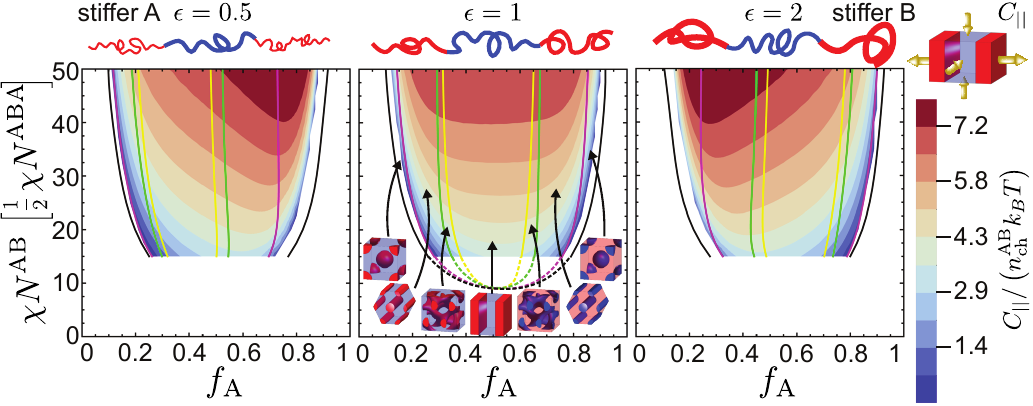}
    \caption{Modulus heat map for 1D lamellar phase for three elastic asymmetries, plotting full region of metastable lamellar phases overlaid by phase diagrams, where the lamellar stability window is between the yellow lines.}
    \label{fig:lam_mod}
\end{figure}

For equilibrium lamellar morphologies, the free energy per chain is only a function of the lamellar spacing $D$; this spacing is determined from the 1D lattice parameter $a$.
Orienting the lamellae along the $z$-axis, the free energy is given by
\begin{equation}
    \Delta F_{\rm mix} = \frac{1}{2}VC_{||}\strain_{zz}^2
\end{equation}
where $C_{||}$ describes the longitudinal rigidity of the lamellae and $\strain_{zz} = \Delta a/a_0$ is the layer strain.
This follows the form of layer elasticity for smectic A liquid crystals, neglecting layer bending~\cite{Ajdari1991,Cohen2000,Thompson2004,Man2015}, which would require implementation of strain-gradient terms~\cite{deGennes1993}.
This free energy has an implicit dependence on additional strain components through the incompressibility constraint, ${\rm tr}\,\bstrain = 0$.

Figure~\ref{fig:lam_mod} shows the longitudinal modulus $C_\parallel$ plotted as a heat map overlaid on the phase diagram for conformationally symmetric ($\epsilon = 1$) and asymmetric ($\epsilon = 0.5, 2$) cases (independent phase diagrams for ABA shown in \SI{Figure S1}), with \SI{Figure S2} showing only those regions where lamellae are the most stable phase.
Focusing first on the conformationally symmetric case ($\epsilon = 1$), we find that the layer stiffness is predominantly determined by $\chi N$ with little variation with A-block fraction $f_{\rm A}$ within the stability window (between the yellow lines).
By contrast, conformational asymmetry introduces significant dependence on $f_{\rm A}$, where the greater volume fraction of the stiffer block is stiffer (i.e., high $f_{\rm A}$ is stiffer for stiffer A block, $\epsilon < 1$~\cite{Milner1994}) \edits{which is expected to be a generic feature of BCP melts~\cite{wei2022_3dp,schoonover2024bridging}}.
This monotonic dependence is can be \emph{approximated} by the rule of mixtures, namely $C_\parallel \sim f_{\rm A} C_\parallel^{\rm A} + (1 - f_{\rm A})C^{\rm B}_\parallel$ for suitably extrapolated values of $C_\parallel^{\rm A}$ and $C_\parallel^{\rm B}$, within the lamellar region of the phase diagram.
It is interesting to note that the stiffest regions are metastable structures for $\epsilon\neq 1$, where there is a greater volume of the stiffer block. 
This has implications for deformation-induced ordering, for example shear-induced alignment could result in greater long-time rigidity materials.

Comparing modulus diagrams for ABA and AB as illustrated in \SI{Figure S3}, we see that the general shape and contours of the modulus heat maps are very similar for both ABA and AB, requiring a narrower focus to elucidate differences.
Taking a slice at $\chi N^{\rm AB} = 30$ allows us to compare trends in modulus and free energy across $f_{\rm A}$ (see \SI{Figure S4}). 
ABA triblocks have a slight asymmetry for $\epsilon=1$, with both phase diagram and modulus heat map being slightly offset to the right (higher $f_{\rm A}$), whereas they are symmetric about $f_{\rm A}=0.5$ for AB diblocks. 
The longitudinal modulus is greater for ABA than AB with respect to $f_{\rm A}$ for all $\epsilon$, with minimal difference for $\epsilon=0.5$ except for the highest $f_{\rm A}$, and increasing difference with increasing $\epsilon$.

Comparing the maximum longitudinal moduli for $\chi N^{\rm AB}=30$, we see that both ABA and AB are stiffer for $\epsilon\neq 1$ by about 10\% compared to $\epsilon=1$. For ABA, the longitudinal modulus at $\epsilon=2$ is greater than that at $\epsilon=0.5$ by about 1\%, and the $f_{\rm A}$ at maximum modulus is further from that at $\epsilon=1$.
These results indicate that longitudinal modulus for lamellae is more strongly impacted by having stiffer bridging chains ($\epsilon=2$) than stiffer matrix.
For AB, the longitudinal modulus is symmetric about $f_{\rm A}=0.5$ for $\epsilon=1$ and mirrored for $\epsilon=0.5$ and 2, respecting the invariance of melt properties under simultaneous exchange of \edits{statistical segment} lengths $\ell_{\rm A} \leftrightarrow \ell_{\rm B}$ and A-block fraction $f_{\rm A} \leftrightarrow f_{\rm B} (=1 - f_{\rm A})$.
Additionally, AB moduli are less than those of ABA when normalized by chain density, indicating that doubling $\chi N$ for ABA compared to AB does not account for the entire difference between the two.
Values of $C_{||,\rm max}$ and their associated $f_{\rm A}$ at each $\epsilon$ for all $\chi N$ are recorded for ABA and AB in \SI{Tables S3-4}, with a graphical representation depicted in \SI{Figure S5}. 
The trends discussed for $\chi N^{\rm AB}=30$ are representative for all $\chi N$, with greater $C_{||,\rm max}$ for greater $\chi N$, located at $f_{\rm A}$ further from 0.5.

\begin{table}[ht]
    \centering
    \begin{tabular}{|c|c|c|c|c|}
        \hline
        & \multicolumn{2}{c|}{ABA ($\chi N=60$)} & \multicolumn{2}{c|}{AB ($\chi N=30$)} \\
        \hline
        $\epsilon$ & $C_{||}$ &$f_{\rm A}$  & $C_{||}$ & $f_{\rm A}$ \\
        \hline
        0.5 & 5.84 & 0.71 & 5.58 & 0.67 \\
        1.0 & 5.29 & 0.53 & 5.07 & 0.50 \\
        2.0 & 5.90 & 0.32 & 5.58 & 0.33 \\
        \hline
    \end{tabular}
    \caption {Maximum modulus, normalized to units of $n_{\rm ch}^{\rm AB}k_BT$, and associated $f_{\rm A}$ for the given $\epsilon$ and $\chi N$ for ABA and AB lamellae.}
    \label{tab:lam_maxMod}
\end{table}

Now consider $C_\parallel$ as a function of $\chi N$ at fixed $f_{\rm A}$ (see \SI{Figure S6}), effectively a vertical slice through Figure~\ref{fig:lam_mod}. 
As discussed in the case of $\chi N^{\rm AB}=30$, ABA and AB are both relatively symmetric across $f_{\rm A}=0.5$ for sufficiently large $\chi N^{\rm AB}$ at $\epsilon=1$, with lines of $C_{||}$ vs $\chi N^{\rm AB}$ collapsing for $f_{\rm A}=0.3,0.5,0.7$.
For $\epsilon\neq 1$, the rule of mixtures is generally followed, where stiffness increases with increasing volume fraction of stiff block.
Directly comparing $\epsilon$ for a given $f_{\rm A}$ shows that the modulus for both ABA and AB remains comparable for all $\chi N^{\rm AB}$ for $f_{\rm A}=0.5$, an interesting result given that $f_{\rm A}=0.5$ is on the edge of the stability window for $\epsilon \neq 1$. 
This indicates that having equal volume fractions of the terminal (A) and bridging (B) blocks counteracts the impact of block stiffness with regard to modulus.
Alternatively, low $f_{\rm A}$ moduli increase from  $\epsilon=0.5$ to 1 to 2, whereas high $f_{\rm A}$ increase from $\epsilon=2$ to 1 to 0.5, although some fluctuations occur for low $\chi N^{\rm AB}$. 
Interestingly, regardless as to data separation method (grouped by $f_{\rm A}$ for a given $\epsilon$ or grouped by $\epsilon$ for a given $f_{\rm A}$), we find consistency with the expectation from strong-segregation theory~\cite{Semenov1985_SovPhysJETP} that the modulus scales with $(\chi N^{\rm AB})^{1/3}$ for high enough segregation strengths.


\subsection{2D: Columnar rigidity}

\begin{figure}[t]
    \centering
    \includegraphics[width=0.8\linewidth]{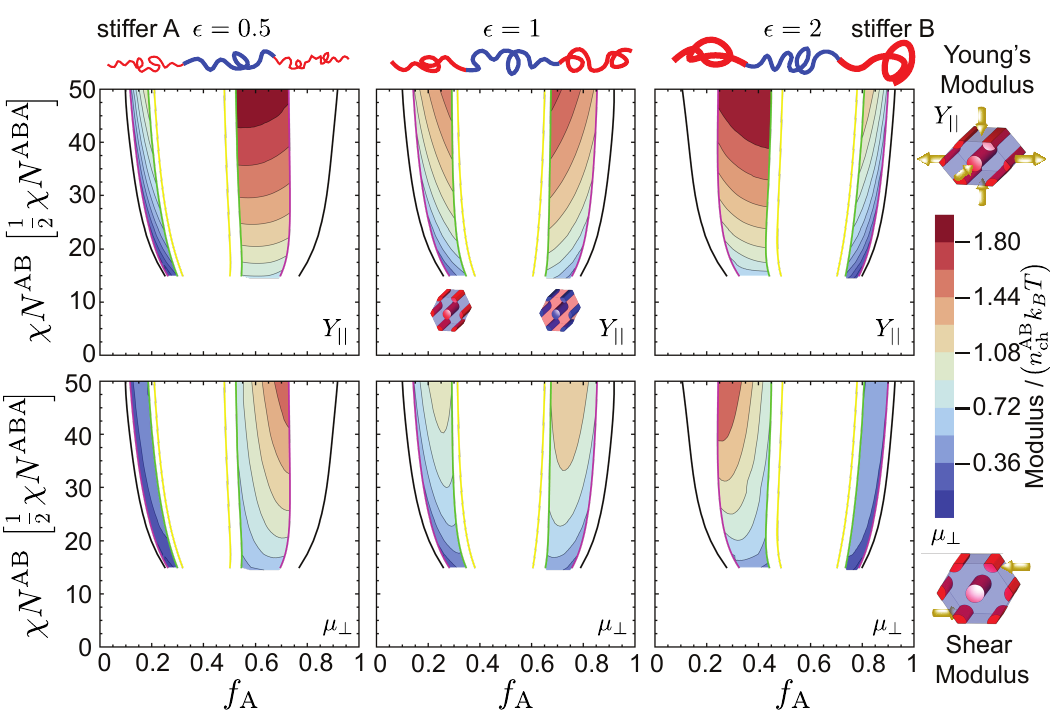}
    \caption{Modulus heat map for 2D hexagonal cylinder phase for three elastic asymmetries for stable phase region, comparing Young's modulus (top) to shear modulus (bottom).}
    \label{fig:hex_mod}
\end{figure}

The cylindrical (columnar) phase has 2D hexagonal symmetry, so is crystal-like in the plane of the cylinders and liquid-like parallel to the cylinders. 
Since the elasticity description of 2D materials with hexagonal symmetry is equivalent to that of isotropic materials, the deformation free energy of mixing is simply
\begin{equation}
    \Delta F_{\rm mix} = \frac{1}{2}V\left[2\mu_{\perp} (\strain_{xx}^2 + \strain_{yy}^2 + 2\strain_{xy}\strain_{yx}) + \lambda_{\perp}(\strain_{xx} + \strain_{yy})^2\right]
\end{equation}
where $\mu_{\perp}$ is the transverse shear modulus and $\lambda_{\perp}$ is the transverse Lam\'e parameter~\cite{LandauLifshitz1986}.
Expressing this in terms of volume-preserving uniaxial deformation along the longitudinal direction, the volume constraint implies that $\strain_{xx} + \strain_{yy} = -\strain_{zz}$ so we can re-express the free energy in terms of a longitudinal Young's modulus $Y_\parallel$  as
\begin{equation}
    \Delta F_{\rm mix} = V\left(\frac{Y_{\parallel}}{2}\strain_{zz}^2 + 2\mu_{\perp} \strain_{xy}\strain_{yx}\right)\, ,
\end{equation}
where $Y_\parallel \equiv \mu_{\perp} + \lambda_{\perp}$, such that under an applied stress $\sigma_{zz}$ along the $z$-axis, we recover the extensional stress-strain relationship $\sigma_{zz} = Y_\parallel \strain_{zz}$.
To determine these parameters, we used two separate deformation protocols applied to a unit cell of the hexagonal cylinder phase.
The first protocol increases the lattice parameter $a$ while fixing $b$ and $\theta (= 2\pi/3)$ to be constant.
The second protocol was an isotropic biaxial stretch of parameters $a$ and $b$, effectively ``swelling'' the in-plane dimensions.
Due to translational symmetry along the $z$-axis (i.e.,~the $c$ axis), while these deformations change the cross-sectional area, the total volume can be preserved by freely re-scaling the height of the 3D unit cell, without altering the free energy per chain.
As shown in \SI{Section S2}, these two protocols provide enough information to determine $Y_\parallel$ and $\mu_\perp$.


Modulus heat maps for Young's and shear modulus ($Y_\parallel$ and $\mu_{\perp}$, respectively) are shown in Figure~\ref{fig:hex_mod}, plotted only in their stable phase regions, with moduli for metastable results shown in \SI{Figure S7}. 
Shear moduli have local maxima with respect to slices at various $\chi N$ within the stability window for $\epsilon=1$, and local maxima on the phase boundary for $\epsilon\neq 1$, as illustrated in \SI{Tables S5-8}. 
As with lamellae, the stiffest regions correspond to greater volume fraction of the stiffer block. 
For 2D and later 3D phases, we introduce the variable of matrix vs filler component; for the columnar phase, the stiffest response is achieved when the stiffer block is the matrix.

Considering both \SI{Figure S8} and \SI{Table S5}, we can see that $f_{\rm A}$ at $\mu_{\perp, \rm max}$ ($f_{\rm A,max}$) is very similar for both ABA and AB, differing by only up to 0.03. 
$f_{\rm A,max}$ occurs at the same or very similar values for $\epsilon=1$ and 2 for B matrix (low $f_{\rm A}$), and the same $f_{\rm A}$ for $\epsilon=0.5$ and 1 for A matrix (high $f_{\rm A}$).
While $Y_{||}$ is greater for ABA than AB for $\chi N^{\rm AB}=30$ for nearly all $f_{\rm A}$ and $\epsilon$ (as was the case for lamellae), $\mu_{\perp}$ is only greater for ABA for A matrix regions (all $\epsilon$) and for $\epsilon=2$ for B matrix (see \SI{Figure S9}). 
This indicates an unexpected ``softness'' of the ABA system relative to the AB system, which we explore further in the Discussion section.


Both shear and Young's moduli are smaller for columnar phase than the extensional modulus for lamellae.
This can be rationalized by the variability of chain orientation that exists for the cylindrical melt: while the lamellar extensional modulus describes the stiffness of the melt when layer spacing is strained, the strain applied to the cylindrical phase involves stretching only subset of chains.
As shown in \SI{Figures S10-11} the lamellar stiffness is approached under biaxial (i.e.,~transverse-isotropic) deformations of the cylinder phase, when all chain orientations are subject to the same strain.
Such effects of directional variation will be revisited in our discussion of polycrystalline averages.


\subsection{3D: Cubic rigidity}
\begin{figure}[t]
    \centering
    \includegraphics[width=0.85\linewidth]{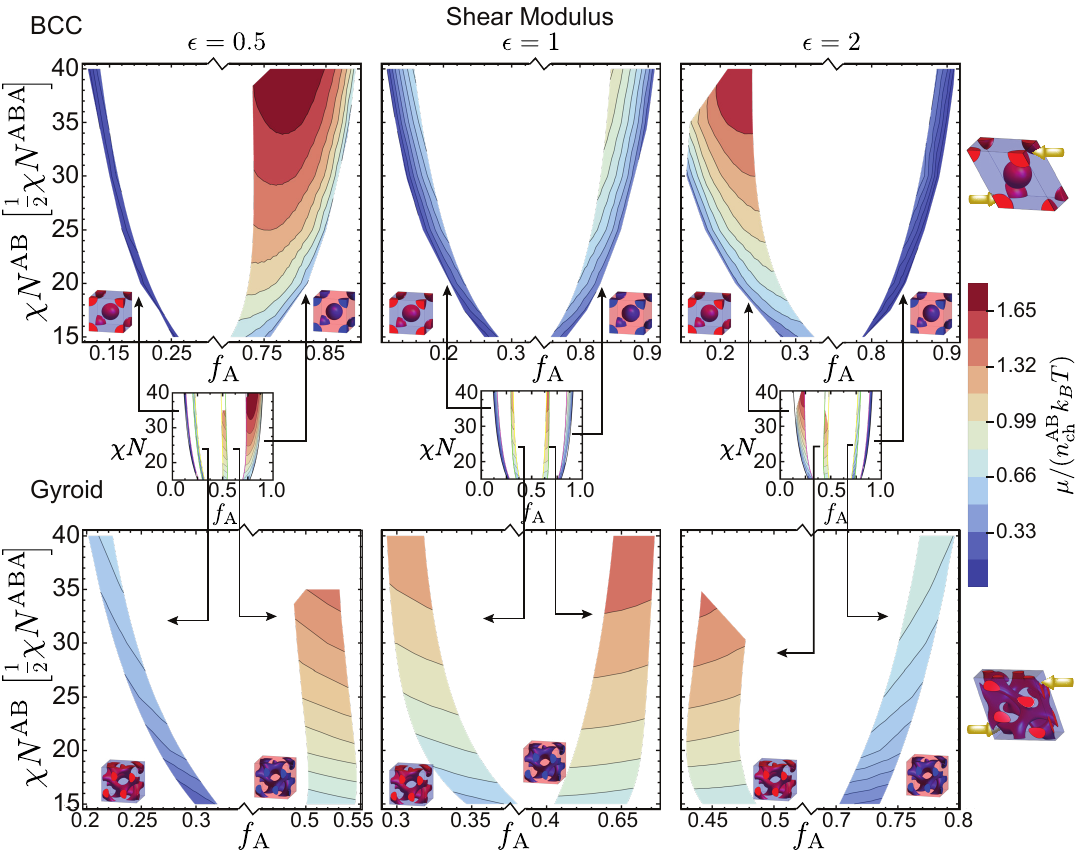}
    \caption{Modulus heat map for 3D cubic phases for three elastic asymmetries for stable phase region, highlighting shear modulus for BCC (top) and gyroid (bottom).}
    \label{fig:cubic_shear_mod}
\end{figure}

The deformation free energy per chain for 3D cubic structures can be written as
\begin{equation}
\begin{split}
    \frac{\Delta F_{\rm mix}}{V} &= \frac{Y}{9}\left((\strain_{xx} - \strain_{yy})^2 + (\strain_{yy} - \strain_{zz})^2 + (\strain_{zz} - \strain_{xx})^2\right) \\
    &\mkern+16mu + 2\mu\left(\strain_{xy}\strain_{yx} + \strain_{yz}\strain_{zy} + \strain_{zx}\strain_{xz}\right)\, ,
\end{split}
\end{equation}
where the bulk modulus is excluded due to the volume constraint (constant volume bulk moduli are undefined).
While this is not an isotropic system, we define $Y$ as an effective Young's modulus, chosen such that under uniaxial stress $\sigma_{zz}$, the stress-strain relation is $\sigma_{zz} = Y\strain_{zz}$; similarly $\mu$ is a shear modulus.
These moduli are calculated using a pair of deformation protocols, where protocol (i) is a volume-preserving uniaxial strain and protocol (ii) is a volume-preserving simple shear; see \SI{Section S1} for more details.
Notably, there are no ``liquid directions'' for cubic structures---these 3D phases obey the rules of crystal elasticity rather than liquid crystal elasticity.
Rather, the liquid response gives rise to non-affine internal rearrangements, captured with a ``liquid network'' description~\cite{Dimitriyev2024} that is distinct from the liquid crystal behavior of the lamellar and columnar phases.
These deformations were performed on $\chi N^{\rm ABA}=30-80$ in increments of 10 with $f_{\rm A}$ spacings of $0.02-0.04$, or $0.01$ for high interest regions (i.e.,~within stable phase windows).
Shear moduli for BCC spheres and double gyroid network phases within their stability windows are plotted in Figure~\ref{fig:cubic_shear_mod} on condensed $f_{\rm A}$ axes, with Young's moduli plotted in \SI{Figure S12}.
Heat maps plotted on the full $f_{\rm A}$ axis can be found in \SI{Figure S13} for a to-scale comparison, and metastable moduli are shown in \SI{Figures S14-15}. 
Note that the Young's modulus is found to be less than the shear modulus ($Y/\mu < 1$) for this system.
For 3D \emph{isotropic} materials $Y/\mu = 2(1 + \nu)$ for Poisson ratio $\nu$, so $Y/\mu < 1$ would imply an auxetic response.
We emphasize that this bound does not hold for our calculations owing to the constant volume (isochoric) conditions under which the moduli are calculated; under these conditions the Poisson ratio is not defined. 


For $\epsilon=1$, gyroid had higher moduli than BCC, and in both cases, the stiffness is larger for the case of an A-block matrix (the ``inverted'' structures with $f_A \gtrsim 0.5$) than the case of a B-block matrix (structures with $f_A \lesssim 0.5$).
Interestingly, whereas the moduli for gyroid have little variation with $f_A$, there is a strong variation of the BCC moduli with $f_A$.
This is likely due to the proximity of the BCC structures to the disordered phase, where the response of the melt to applied deformations can destroy the ordered phase.
Indeed, we find that the modulus drops rapidly for all phases as their composition reaches the edge of the range for which ordered solutions to the SCFT calculations were found, hinting at loss of stability of the ordered phases.
This is particularly true for all $Y$, and for $\mu$ for $\epsilon=1$ and the smaller portion of phase diagram for $\epsilon\neq1$ (B matrix for $\epsilon=0.5$, A matrix for $\epsilon=2$).
For BCC, the values of $\mu$ for $\epsilon\neq1$ find distinct maxima within the stable phase region whenever the matrix block is stiffer than the minority block (see \SI{Tables S9-11} and \SI{Figure S16}); similar trends can be seen in the shear modulus of the cylinder phase (i.e.,~the shape of the $\mu$ heat map is comparable for both columnar and cubic phases).
The maxima for gyroid $\mu$ are less well defined due to the narrow phase stability window, although when extending the calculations into metastable regions, we were able to locate distinct maxima for each $\chi N$ (see \SI{Tables S12-14} and \SI{Figure S17}).
Note that our calculations for the gyroid phase encountered convergence issues for $\epsilon \neq 1$, particularly for structures with a stiffer matrix block.

As with the columnar phase, ABA moduli show similar trends as AB moduli, although there is a slight asymmetry where the moduli are shifted to slightly higher values with increasing $f_{\rm A}$.
For both gyroid and BCC, Young's moduli are greater for ABA than AB at $\chi N^{\rm AB}=30$, and shear moduli are greater for A matrix for all $\epsilon$ and also B matrix for $\epsilon=2$ (see \SI{Figures S18-19}).
Interestingly, the magnitude of the moduli is on a similar scale to columnar phases, and about 1/4 that of lamellae.
Since cubic structures are expected to be the stiffest, this demonstrates that ideal structures without defects or different orientations produce different results than intuitive expectations. 
This can be adjusted for by taking polycrystalline averages, as discussed below.

\section{Discussion}

\subsection{Rule of mixtures for 1D lamellae}

The mechanics of BCP melts are often understood in the context of composites: a rubbery block and a glassy block respectively correspond to the soft and hard components of a composite.
Accordingly, it is expected that the macroscopic stiffness of the material can be represented as an average over the stiffnesses of the component blocks.
In the case of a 1D lamellar phase, the layers are stretched in series, and thus one might consider that the layers are under an isostress condition.
However, due to the local volume balance constraint, in straining the interlayer spacing from $D$ to $D'$, the A-block subdomain is strained from $f_{\rm A} D$ to $f_{\rm A} D'$, so the B-block subdomain must be strained from $(1-f_{\rm A}) D$ to $(1-f_{\rm A}) D'$.
Therefore, the lamellar phase is in an isostrain condition under longitudinal deformations, and thus it is expected that the longitudinal modulus obey the rule of mixtures, namely $C_\parallel \approx C_{\rm A}\,f_{\rm A} + C_{\rm B} (1 - f_{\rm B})$, where $C_{\rm A}$ and $C_{\rm B}$ are the pure-A block and pure-B block moduli, respectively.

In order to approximate the single-block stiffnesses $C_\alpha$ (for $\alpha = {\rm A,B}$), we turn to strong-segregation theory (SST) for AB diblocks, which is the asymptotic limit of SCFT when $\chi N \to \infty$~\cite{Semenov1985_SovPhysJETP}.
Here, the IMDS is taken to be infinitely narrow, so that the enthalpic cost of mixing is $H_{\rm mix} = \gamma A = 2\gamma V/D$, where $A$ is the IMDS area and is a surface free energy, and $V = AD$ is the unit cell volume.
The single-block subdomains are then approximated as polymer brushes and the stretching free energy cost is computed using parabolic brush theory, yielding $F_{{\rm str.},\alpha} = (\pi^2/32)n_{\rm ch} k_B T D^2 f_\alpha^3/(N_\alpha \ell_\alpha^2)$, where $N_\alpha$ is the degree of polymerization and $\ell_\alpha$ is the \edits{statistical segment} length of block $\alpha$.
As shown in \SI{Section S4}, the longitudinal modulus can then be expressed as
\begin{equation}\label{eq:SST_C1}
    C^{\rm SST}_\parallel = \frac{4\gamma}{D} + 2\kappa\frac{D^2}{N\ell_{\rm A}^2}f_{\rm A} + 2\kappa\frac{D^2}{N\ell_{\rm B}^2}\left(1-f_{\rm A}\right)\, ,
\end{equation}
where $\kappa \equiv (\pi^2/32)n_{\rm ch}k_B T$.
Note that the last two terms have the form of a rule of mixtures, where the entropic stiffness of a polymer with degree of polymerization $N$ and \edits{statistical segment} length $\ell_\alpha$ is given by $\kappa D^2/(N\ell_\alpha^2)$.
While the entropic stiffness of Gaussian chains that appears in Flory's theory of rubber elasticity scales with $R^2/(N \ell^2)$, where $R$ is the separation length of chain ends, parabolic brush theory allows for a variety of chain ends to be sampled, so that this length is re-expressed as the domain spacing $D$, along with a numerical prefactor.
Importantly, the stiffness of microphase separated domains acquires an additional dependence on the effective enthalpy of mixing, as represented by the first term.
This describes how volume-preserving deformations that extend the domain spacing $D$ must also decrease the area $A$ of the IMDS.
As the interface is always assumed to exist within SST, regardless of $f_{\rm A}$, this term falls outside of the rule of mixtures.

For comparison with the SCFT results where the domain spacing $D$ is unconstrained, chosen such that the free energy of mixing is minimized, we solve for the equilibrium spacing, yielding a new expression for the equilibrium longitudinal stiffness,
\begin{equation}\begin{split}
    C^{\rm SST}_{||}\big|_{\rm D^*} &= \kappa^* \times\left[\sqrt{N \ell_{\rm A}^2}\sqrt{N \ell_{\rm B}^2} \left(\frac{f_{\rm A}}{N\ell_{\rm A}^2}+\frac{1-f_{\rm A}}{N\ell_{\rm B}^2}\right)\right]^{1/3}\\
    &= \kappa^* \times\left(\frac{f_{\rm A}}{\epsilon}+\epsilon(1-f_{\rm A})\right)^{1/3}\, ,
\end{split}\end{equation}
where $\kappa^* \equiv \left(\frac{9\pi^2}{8}\right)^{1/3} n_{\rm ch}k_BT (\chi N)^{1/3}$.
Note that the rule of mixtures, based on a simple arithmetic average of block stiffnesses, is absent in its usual form. 
Instead, there is a weighting factor that depends on the geometric average of \edits{statistical segment} lengths, arising from the enthalpic cost of interaction between chains~\cite{Helfand1975,Helfand1975_2}.
This combination acquires a nonlinearity---the cube root---that accounts for the effect of changes in equilibrium spacing based on the balance of interfacial enthalpy and entropic costs of chain stretching. 
This new average can be expressed completely in terms of $f_{\rm A}$ and the conformational asymmetry $\epsilon$.
Despite the loss of the traditional rule of mixtures, we still find a monotonic dependence of the melt stiffness with $f_{\rm A}$ at fixed $\epsilon$.
However, the nonlinearity of this relationship points to parameter regions where relatively large changes in melt stiffness can be achieved with small changes in $f_{\rm A}$; this is particularly true for cases of large conformational asymmetry $\epsilon \gg 1$ or $1/\epsilon \gg 1$.
Importantly, the stiffness of the melt is expected to be constant with $f_{\rm A}$ for $\epsilon = 1$, which is consistent with SCFT measurements within the stability range for lamellae; the SST predictions fail as the phase loses stability or approaches the order/disorder transition.
While this result was derived for AB diblock copolymer melts, the key takeaways should hold for ABA melts as well\edits{, although as we note in the next section, there are important differences arising from finite-segregation corrections}.\remove{; as we show below, however, the presence of bridging chains can dramatically alter melt stiffness, particularly in non-lamellar melts.}
Finally, note that the overall prefactor $\kappa^*$ accounts for the usual scaling with chain density and temperature, along with a factor of $(\chi N)^{1/3}$; as shown in \SI{Figure S6}, our SCFT results are consistent with this scaling for high enough segregation strengths.

\subsection{Domain spacing \edits{and segregation strength}}

\begin{figure}[t]
    \centering
    \includegraphics[width=0.37\linewidth]{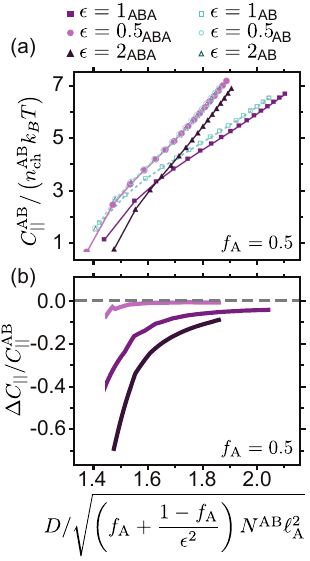}
    \caption{(a) Plot showing the scaling of longitudinal modulus with domain spacing for lamellae at $f_{\rm A}=0.5$, with domain spacing normalized by $\epsilon$ and $f_{\rm A}$. (b) Plot showing the percent difference between $C_{||}$ for ABA and AB, where $\Delta C_{||} = C_{||}^{\rm ABA}-C_{||}^{\rm AB}$.}
    \label{fig:lam_Dspacing}
\end{figure}

It is widely accepted that the $D$-spacing of a BCP melt plays a key role in determining the melt's stiffness~\cite{Bates1999}.
In general, ABA and AB melts can have different $D$-spacings at equivalent molecular parameters and segregation strengths.
To assess the degree to which additional contributions arising from polymer architecture may affect melt stiffness, we compare ABA and AB melts as a function of $D$.
The $D$-spacing is determined in units of the RMS radius of the polymer coil, $\sqrt{\left(f_{\rm A} + (1-f_{\rm A})/\epsilon^2\right)N^{\rm AB}\ell_{\rm A}^2}$.
The longitudinal modulus for lamellae is plotted with respect to normalized $D$-spacing for $f_{\rm A}=0.5$ (Figure~\ref{fig:lam_Dspacing}(a), see \SI{Figure S20} for $f_{\rm A}=0.3,0.7$).
We find that the longitudinal modulus monotonically increases with $D$ for the range of $\chi N$ values sampled, consistent with expectations of eq.~\ref{eq:SST_C1}.
Interestingly, the longitudinal modulus is consistently greater for AB than ABA with respect to $D$ for all $\epsilon$ as clearly demonstrated in Figure~\ref{fig:lam_Dspacing}(b).
Our results show that the stiffness comparisons between ABA and AB melts require care: if compared at fixed $\chi N$, ABA melts are stiffer, whereas AB melts are stiffer when compared at fixed $D$-spacing.
\edits{
However, as shown in SI figure S24, the $D$-spacing for ABA melts is somewhat larger than that of AB melts at equivalent $\chi N$; this has been confirmed in experiments~\cite{Mai2000}.
Therefore, in order to fix the $D$-spacing between ABA and AB melts for comparison, a somewhat smaller value of $\chi N$ was chosen for ABA melts, depressing the calculated stiffness.
Since the most practical way to tune $D$-spacing in experiments is through the degree of polymerization $N$, it is prudent to consider the accompanying change in segregation strength when comparing stiffnesses of different polymer architectures, rather than $D$-spacing alone.
}
The key difference between ABA and AB melts is the presence of the middle B block, which we can regard as a cross-linking site between the ends of two B blocks for a pair of AB chains.
This added constraint should affect chain packing statistics and thus melt thermodynamics.
This is confirmed by the fact that the difference in melt stiffness is greatest for $\epsilon > 1$, when the B block is stiffer than the A block, magnifying differences in chain stretching costs.
Meanwhile, the difference is minimized for $\epsilon < 1$, when the A block is stiffer; since the A block is identical for ABA and AB systems, the difference in comparative melt stiffness approaches zero.
Further, note that the difference in melt stiffness seems to approach zero as the $D$-spacing becomes large. 
\edits{
Given that finite-segregation to strong-segregation theory are dominated by the entropy of chain ends, it is reasonable to expect that differences in ABA and AB melt stiffnesses decrease with increasing $\chi N$~\cite{Likhtman2000,Matsen2010}.
This difference is further evidenced by the difference in the critical point $(\chi N)_{\rm ODT}$ between AB and ABA copolymers~\cite{Mayes1989,Mayes1991,Gehlsen1992}.
}


\subsection{Polycrystalline Averages}

\begin{figure}[t]
    \centering
    \includegraphics[width=\textwidth]{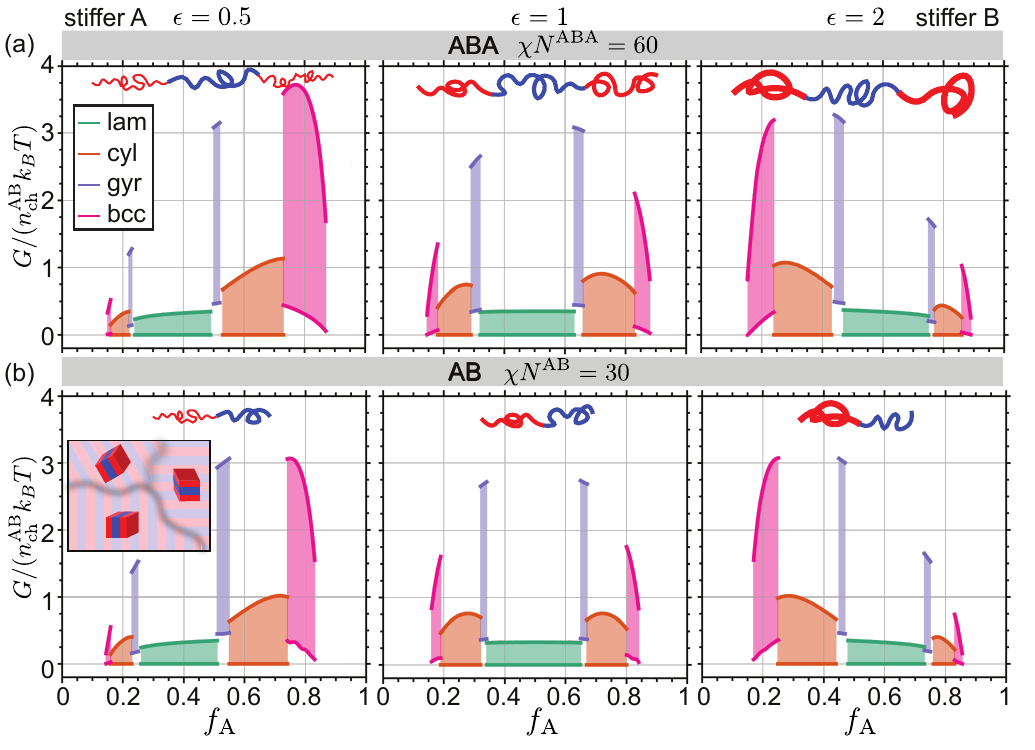}
    \caption{Comparison of polycrystalline average modulus bounds (upper bound $G_V$, lower bound $G_R$) for all phases for (a) ABA and (b) AB melts at $\chi N^{\rm AB}$=30.}
    \label{fig:poly_avg_mod}
\end{figure}

While we report the anisotropic elastic moduli for perfect grains of each of the canonical equilibrium linear BCP phases, real materials are polycrystalline, containing multiple grain orientations, as well as defects such as dislocations and grain boundaries~\cite{Read1999,Knoll2002,Jinnai2006,Mareau2007,Kim2010,Pezzutti2011,Ryu2013,Li2015,Feng2021,Feng2023}.
The \edits{equilibrium} mechanics of real systems ultimately depends on the statistics of grain orientation and defect distributions, which in turn depends on material processing.
Moreover, the scale of the SCFT computations required to capture these behaviors is prohibitive, although progress has been made in the cases of lamellar and cylindrical phases~\cite{Nagpal2012,Li2016,Blagojevic2023}.
Nonetheless, we can obtain upper and lower bound estimates of polycrystalline melt stiffness, based on averages of anisotropic moduli over a distribution of grain orientations.
Since a uniform distribution of grain orientations is assumed, there are no symmetry assumptions for the resulting polycrystal, so it is expected to behave like a 3D isotropic elastic medium, and is therefore characterized in terms of two elastic moduli.
As one of these moduli can be taken to be the bulk modulus, which is undefined for our volume-constrained calculations, it is sufficient to determine the macroscopic shear modulus $G$.
Treating the lamellar and cylindrical phases as lower-dimensional slices of 3D materials with an orthorhombic unit cell, we calculated the Voigt $G_V$ and Reuss $G_R$ averages of the macroscopic shear modulus~\cite{berryman_bounds_2011} (see \SI{Section S5} for details) and plotted the results in their stable regions for fixed $\chi N$ in Figure~\ref{fig:poly_avg_mod} (for a full heat map across $\chi N$, see \SI{Figure S21}).
\edits{Importantly, these bounds should be considered as the \emph{limit} of the polycrystalline average as certain components of the elastic modulus tensor are taken to zero; see the SI discussion for more details.}

Comparing the upper polycrystalline average to perfectly-ordered structures, we note that the lamellar moduli were drastically reduced and relatively flattened, whereas columnar moduli were relatively consistent and cubic were increased.
Notably, the Reuss bound is zero for lamellae and columnar phases, as they have liquid-like directions that dominate the response under isostress conditions.
This suggests that while all structured phases display \edits{non-equilibrium} behavior, 1D and 2D ultimately fail to be rigid due to an almost-certain presence of a grain orientation that allows for liquid-like flow.
However, Reuss bounds for cubic phases are nonzero, suggesting a plateau behavior at terminal times (i.e.,~zero-frequency response, $\omega\to 0$).

\edits{
Comparing the calculated shear moduli to literature results, we find that experimental values generally fall within the calculated bounds across a range of polymers and ordered phases for both diblock and triblock systems.
Experimental comparisons were chosen from literature that reported shear moduli approaching order-order or order-disorder transitions (generally collected at low frequency) and provided sufficient polymeric parameters to enable comparison to calculated values.
In one study, Almdal et al.~reported the moduli of PI-PDMS and poly(ethylene-alt-propylene) (PEP)-PDMS diblocks in lamellar and gyroid phases approaching their order-disorder transitions~\cite{almdal1996order}.
Their moduli measured just below the order-disorder temperature ($T_{\rm ODT}$) fall within our calculated modulus upper and lower bounds as shown in SI Table S15, converted to kPa using the experimental temperature, degree of polymerization (accounting for composition fluctuations), and reference volume (118 \AA$^3$) reported for each diblock.
Likewise, moduli for PS-PI diblocks and PS-PI-PS triblocks forming both hexagonal cylinders and BCC spheres fell within calculated bounds, indicating a level of reproducibility across polymeric platforms~\cite{ryu1997structure}.
Notably, the experimental and calculated parameters ($\chi N$, $\epsilon$) do not align in all cases, but the calculated bounds provide a sufficient benchmark for comparison.
Also worth noting, the experimental moduli generally lie closer to the lower bound moduli, suggesting significant relaxation of domains along their liquid-like directions or the presence of defects, which is not captured within the polycrystalline averages.
Additionally, experimental moduli are greater than calculated for crystalline samples, and lower for entangled samples~\cite{floudas2000nucleation,Kossuth1999}.
}

Experimental testing on ceramic composites demonstrates that gyroid is inherently mechanically superior to BCC~\cite{Sokollu2022}, and it is notable our results do agree with this despite the long timescales assumed for our computations (with the exception of a range of high-$f_{\rm A}$ structures when $\epsilon=0.5$).
However, in the case of BCPs, this may also be partly due to greater volume fraction asymmetry for BCC compared to gyroid: approaching the disordered (``miscible'') phase lowers the stiffness of the material, apart from the inherent mechanical properties of the morphology itself, since the \edits{equilibrium} rigidity relies on the existence of microphase separated domains.
Indeed, the composite stiffness, while dependent on morphology, adopts an overall trend that conforms to the expected BCP miscibility gap.
We note that in all cases, cubic is stiffer than 2D is stiffer than 1D structures, in agreement with Kossuth, et al.~\cite{Kossuth1999} (\textit{viz.}, Figure~\ref{fig:scheme_def_overview}(a)).

As hinted in Figure~\ref{fig:poly_avg_mod}, the Voigt (upper) bounds for the 3D phases form a rather smooth ``envelope'' that is concave down, centered around $f_{\rm A}=0.55$ for $\epsilon=1$, and shifted to higher and lower $f_{\rm A}$ for $\epsilon=0.5,2$ respectively.
This result is born out in heat maps of Voigt (upper) modulus bounds for each $\epsilon$ shown in \SI{Figure S21}: level sets seemingly match across phase boundaries, albeit with drastically different absolute values of modulus.
Taking the convex hull of gyroid and BCC Voigt bounds gives a smooth curve as shown in \SI{Figure S22}, indicating a characteristic upper bound modulus for 3D BCP phases that is not expected to be particularly sensitive to the underlying morphology, although we have not confirmed that the envelope accurately predicts the stiffness of competing 3D phases.
Nevertheless, we propose that this convex hull would include any long-lived response of metastable 3D phases, such as the double diamond network phase or more exotic sphere phases.

\begin{figure}[t]
    \centering
    \includegraphics[width=0.8\linewidth]{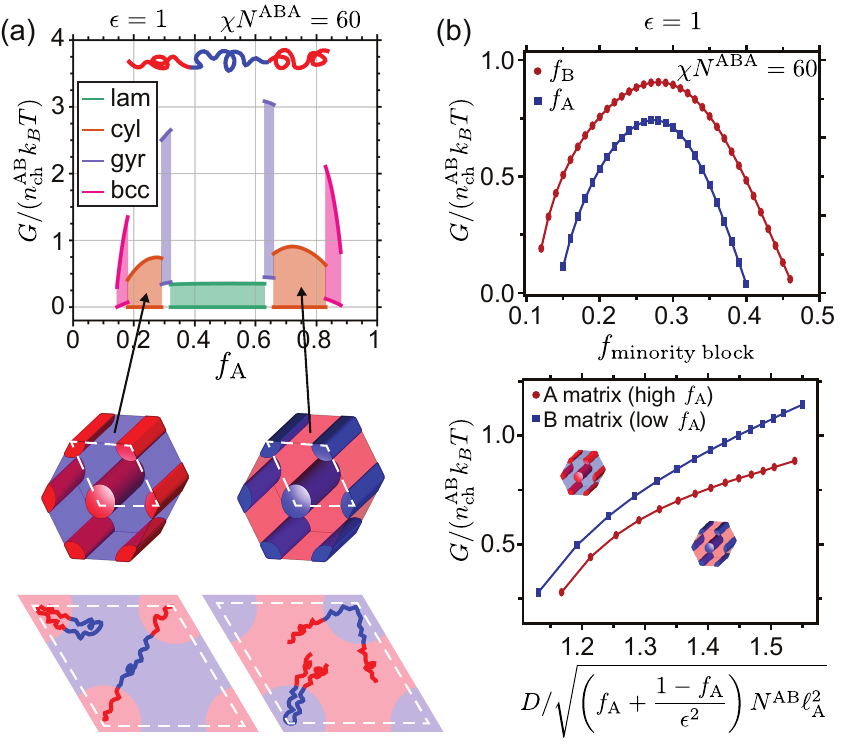}
    \caption{Modulus asymmetry for ABA triblocks, as illustrated in the case of columnar phase for $\chi N^{\rm ABA}=60$, $\epsilon=1$. (a) Modulus plot for all phases, illustrating the difference between B matrix (left) with bridging and looping chains vs A matrix (right) without bridging chains. (b, top) Modulus for columnar phase plotted against $f_{\rm minority\:block}$ with minority A as blue squares and minority B as red circles. (b, bottom) Modulus plotted against domain spacing, $D$, normalized by $f_{\rm A}$ and $\epsilon$, separated into A matrix ($f_{\rm A}=0.7$, red circles) and B matrix ($f_{\rm A}=0.3$, blue squares).}
    \label{fig:mod_asym}
\end{figure}

As discussed in the case of both columnar and cubic rigidity, for ABA melts, larger $f_{\rm A}$ structures are stiffer for $\epsilon=1$ and equivalent $f_{\rm A}$ for $\epsilon=0.5,2$.
In particular if we study the case of $\epsilon=1$, it is counterintuitive that A matrix melts are stiffer than B matrix since there are no bridging chains in that case.
Indeed, bridging chains are commonly used to stiffen and toughen thermoplastics,~\cite{Honeker1996Impact,Eagan2017Combining} and yet the stiffer melt is achieved in the inverted phases, in which there are no bridging chains, as illustrated in Figure~\ref{fig:mod_asym}(a).
In order to further probe this unintuitive result, we compared the modulus for each conformation with respect to domain spacing $D$ (tuned via segregation strength $\chi N$), not just $f_{\rm A}$, shown in Figure~\ref{fig:mod_asym}(b). 
This comparison revealed that bridging conformations have greater modulus with respect to $D$ than non-bridging conformations, further demonstrating the importance of $D$ when comparing moduli of ``equivalent'' systems.
Thus, complementary methods of designing melts---direct control over minority domain block fraction or indirect control over $D$ (e.g.~via segregation strength $\chi N$)---alters the projected response of the melt.

To confirm that the modulus calculations involve bridging chains, we use the method described by Matsen and Thompson~\cite{Matsen1999} (see \SI{Section S6} for more details) to compute chain bridging fractions, finding that the ``normal'' low-$f_{\rm A}$ cylinder phase consistently has bridging fractions in the range of $60-75\%$, as shown in \SI{Figure S23(b)}.
\remove{As we discuss below, since the terminal rigidity quantifies the ability for melts to undergo long-time relaxation under fixed strain, we can attribute this unexpected softening effect of bridging chains to an additional structural relaxation that they enable.}
\edits{
Moreover, there is little change in the bridging fraction of either lamellae or cylinder phases under 10\% strain.
We find that lamellae maintain a bridging fraction $\nu_b \approx 48-52\%$ both before and after deformation over a range of compositions and conditions, consistent with theoretical predictions of strong-segregation theory (see \SI{Figure S23(a)})~\cite{Milner1992,Semenov1995,Bjorling1998_1,Bjorling1998_2}.
In fact, for any single melt, we find that the change in bridging chains $\Delta \nu_b$ is small, typically $\Delta \nu_b \lesssim 0.5\%$ for 10\% strain.
The change in bridging fraction for a columnar system is somewhat larger, often on the order of a few percent.
It has recently been shown~\cite{White2025} that there is significant exchange of chains between domains over timescales that exceed the typical macroscopic stress relaxation time.}

\subsection{\edits{Bending rigidity of lamellae and cylinders}}
\edits{
While lamellae and cylinders lack rigidity along their ``liquid'' directions, consistent with their identification as smectic-A and columnar phases, a mapping to liquid crystal elasticity requires incorporation of orientational elasticity.
While a full characterization of the Frank free energy is beyond the scope of this work, the bending stiffness $K_b$ of each phase provides a key measure of the scale over which orientational elasticity becomes important.
Using a small deflection approximation~\cite{deGennes1993}, the bending free energy of either lamellar or cylinder phases can be approximated as}
\edits{
\begin{equation}\label{eq:bending}
F_{\rm bend} \approx \frac{1}{2}VK_{\rm bend} \int_0^\lambda\frac{{\rm d}z}{\lambda}\left(\frac{\partial^2 u_\perp}{\partial z^2}\right)^2
\end{equation}
}
\edits{where $u_\perp$ is the deflection of the middle axis of a domain transverse to the direction of translational symmetry, which is taken to be the $z$ axis.
This deflection is assumed to be periodic with wavelength $\lambda$.
Furthermore, a characteristic \emph{bending length} $\xi_{\rm bend}$ can be determined from an elastic modulus $G$ through the ratio $\xi_{\rm bend} \equiv \sqrt{K_{\rm bend}/G}$.
A similar lengthscale was derived in the context of molten polymer brushes with rippled surfaces, which is relevant for the strong-segregation limit of block copolymer melts~\cite{Fredrickson1992}.}

\edits{Unlike the homogeneous strain imposed by controlled changes of lattice parameters, direct control over strain gradients is difficult for computational SCFT.
To accomplish this, we turn to the use of external mean fields $h_\alpha$ that couple to the volume fraction field $\phi_\alpha$, where $\alpha \in \{{\rm A},{\rm B}\}$ denotes the monomer index.
External fields have recently been used to simulate thin film confinement and other boundary effects~\cite{Magruder2025}.
The external field directly modifies the chemical potential field $w_\alpha$ to the form}
\edits{
\begin{equation}\begin{split}
w_{\rm A}^\eta(\mathbf{r}) &= \chi\phi^\eta_{\rm B}(\mathbf{r}) + \Xi(\mathbf{r}) + \eta h_{\rm A}(\mathbf{r})\,\,\, \text{and} \\
w_{\rm B}^\eta(\mathbf{r}) &= \chi\phi^\eta_{\rm A}(\mathbf{r}) + \Xi(\mathbf{r}) + \eta h_{\rm B}(\mathbf{r})\, ,
\end{split}\end{equation}
}
\edits{where $\Xi$ is a Lagrange multiplier field used to maintain the incompressibility constraint $\phi_{\rm A}(\mathbf{r}) + \phi_{\rm B}(\mathbf{r}) = 1$ and $\eta$ is a perturbation parameter that characterizes the strength of the external field $h_\alpha$.
Note that with the introduction of the external field, the volume fraction field of the equilibrium melt is perturbed from its $\eta = 0$ configuration.
The resulting free energy per chain functional $\tilde{F}'_\eta$ then has the form}
\edits{
\begin{equation}
\tilde{F}'_\eta = \tilde{F}_{\rm F-H}[w_\alpha^\eta, \phi_\alpha^\eta] + \frac{\eta}{V}\int{\rm d}V_r\left[h_{\rm A}(\mathbf{r})\phi^\eta_{\rm A}(\mathbf{r}) + h_{\rm B}(\mathbf{r})\phi^\eta_{\rm B}(\mathbf{r})\right]\, ,
\end{equation}
}
\edits{where $\tilde{F}_{\rm F-H}$ is the standard ``Flory-Huggins'' free energy per chain of the melt without the external field contribution and is composed of the ideal chain and two-body interactions.
Given an equilibrium configuration of the melt under the influence of the external field, the Flory-Huggins free energy can be represented as a function of the perturbation strength $\eta$ as $\tilde{F}_{\rm F-H}(\eta) \equiv \tilde{F}_{\rm F-H}[w_\alpha^\eta, \phi_\alpha^\eta]$; assuming that $h_\alpha$ breaks the continuous translational symmetry of a stable lamellar or cylindrical phase, the Flory-Huggins free energy must increase as $\tilde{F}_{\rm F-H}(\eta) \approx \tilde{F}_{\rm F-H}^0 + \frac{1}{2}K_h\eta^2$ with $K_h >0$ for small perturbations ($\eta \ll 1$).

In order to determine the strain gradient in eq.~\ref{eq:bending}, we must define a suitable displacement field $u_\perp$ describing the deflection of interfaces within the melt, as shown in Fig.~\ref{fig:8-new}.
At fixed values of $z$, we can define the center of mass of either block as the integral
\begin{equation}
\mathbf{R}^{\eta}_\alpha(z) \equiv \frac{\iint{\rm d}x\,{\rm d}y\, \mathbf{r}\,\phi^\eta_\alpha(\mathbf{r})}{\iint{\rm d}x\,{\rm d}y\, \phi^\eta_\alpha(\mathbf{r})}\, ,
\end{equation}
where integrals are taken over $xy$-plane slices along the $z$-axis.
The transverse displacement can then be found from the difference $u^\eta_\perp(z) \equiv |\mathbf{R}^\eta_\alpha(z) - \mathbf{R}^0_\alpha(z)|$ where $\alpha$ is chosen to correspond to the minority block.
Taking $u^\eta_\perp$ to be a linear response to a small applied field, ${\rm d}^2u^\eta_\perp/{\rm d}z^2 \sim \eta$, so that the bending integral in eq.~\ref{eq:bending} and $\tilde{F}_{\rm F-H}(\eta)$ both scale with $\eta^2$ for $\eta \ll 1$.
Therefore, by iteratively increasing $\eta$ and measuring both the bending integral and the change in free energy per chain, the bending modulus can be determined by a simple linear fit.

\begin{figure}[t]
\includegraphics[width=0.6\textwidth]{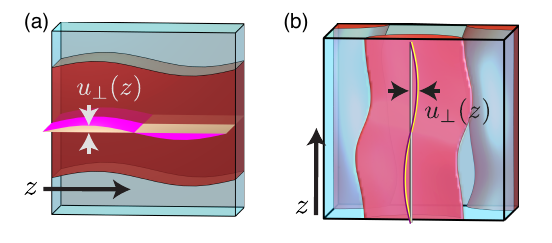}
\caption{\label{fig:8-new} Depiction of bent lamellar (a) and cylinder (b) melts, along with displaced centers of mass (magenta and yellow, respectively) for the A-block domain. Also shown: the transverse displacement field $u_\perp(z)$ describing the deflection of the center of mass from its un-bent configuration.}
\end{figure}

For the lamellar case, we focus on $f_{\rm A} = 0.4$ and $0.6$ for both AB and ABA systems, choosing $\chi N^{\rm AB} = 30$ and $\epsilon = 1$.
For the AB system, we find that $K_{\rm bend} \approx 0.16\times k_BTn^{\rm AB}_{\rm ch}N^{\rm AB}\ell^2$; the longitudinal modulus is approximately $5\times k_B T n^{\rm AB}_{\rm ch}$, resulting in a bending length of $\xi_{\rm bend} \approx 0.18 \sqrt{N^{\rm AB}}\ell$.
For the ABA system, the bending modulus is somewhat larger and has a slightly asymmetric $f_{\rm A}$-dependence, but varies little from $K_{\rm bend} \approx 0.21\times k_BTn^{\rm AB}_{\rm ch}N^{\rm AB}\ell^2$ for the $f_{\rm A}$ values calculated; with a longitudinal modulus of approximately $5.2\times k_BT n^{\rm AB}_{\rm ch}$ the resulting bending length about $\xi_{\rm bend} \approx 0.20\sqrt{N^{\rm AB}}\ell$.
Consequently, curved sections of lamellae can be localized to small lengths, roughly $20\%$ of the r.m.s.~coil size.
This indicates a capacity of lamellar phases to ``heal'' disruptions in structure that lead to bending over relatively short distances and may rationalize the appearance of sharp kinks or chevron structures seen in experiments~\cite{Cohen2000}.

By comparison, the bending modulus for cylindrical structures is much larger.
Choosing $f_{\rm A} = 0.3$ and 0.7 (with 0.7 adopting an inverse, A-matrix structure), we first find that for AB melts, the bending modulus is $K_{\rm bend} \approx 3.84\times k_BTn^{\rm AB}_{\rm ch}N^{\rm AB}\ell^2$; comparing with a Young's modulus of approximately $1.15\times k_BT n^{\rm AB}_{\rm ch}$, we find a bending length of $\xi_{\rm bend} \approx 1.83\sqrt{N^{\rm AB}}\ell$.
For ABA melts, the difference is much more extreme, with $K_{\rm bend}(f_{\rm A} = 0.3) \approx 14.27\times k_BTn^{\rm AB}_{\rm ch}N^{\rm AB}\ell^2$ and $15.94\times k_BTn^{\rm AB}_{\rm ch}N^{\rm AB}\ell^2$; choosing a Young's modulus of $1.18\times k_BT n^{\rm AB}_{\rm ch}$ for $f_{\rm A} = 0.3$ and $1.3\times k_BT n^{\rm AB}_{\rm ch}$ for $f_{\rm A} = 0.7$, the bending lengths are $\xi_{\rm bend} \approx 3.48\sqrt{N^{\rm AB}}\ell$ and $\xi_{\rm bend} \approx 2.14\sqrt{N^{\rm AB}}\ell$, respectively.
While bending deformations of lamellae involve tilt and splay of chains along one dimension, there are more significant disruptions to the polymer packing environments for cylinders, including the polymers that are primarily oriented perpendicular to the bending direction.
Consequently, we can expect that bent cylinders involve a greater degree of packing frustration than bent lamellae, resulting in a greater bending stiffness.
Similarly, the additional constraints imposed by ABA chain architecture should compound the packing frustration of the bent cylinders; indeed, there is a significant increase in bending stiffness for ABA chains.
The corresponding increase in bending length is expected, with ABA cylinders having the largest bending lengths.
Notably, the significant difference in bending length between $f_{\rm A} = 0.3$ and 0.7 ABA systems hints at the important role of bridging chains in extending the range over which bending deformations are expected to be correlated.
}

\section{Conclusions and Outlook}

We have demonstrated that self-consistent field theory (SCFT) can be used to probe the \edits{equilibrium} mechanics of ABA and AB BCP melts, rapidly iterating over multiple configurations and segregation strengths.
Specifically, SCFT can reproduce certain expected results for BCP stiffness, in particular for long time scales: (i) lamellae and columnar phases behave as liquid crystals, with a vanishing lower bound elastic modulus $G_R$; (ii) the upper bound elastic modulus $G_V$ increases in the order of pattern dimensionality, from 1D lamellae to 2D cylinders to 3D cubic phases.
We find that \edits{equilibrium} stiffness comparisons between ABA and AB melts as well as normal (bridging) vs inverse (non-bridging) phases are complicated by preferred representation of information.
The materials structure representation of stiffness as a function of $D$-spacing---often used for the design of thermoplastics~\cite{Bates1999,Mamodia2007Effect}---shows that AB melts are stiffer than ABA melts and inverse phases of ABA melts are less stiff than the normal phases; the synthetic chemistry representation as a function of $f_{\rm A}$ \emph{at fixed $\chi N$} reports opposite results.
\remove{In particular, the observed reduction in stiffness due to bridging chains at fixed $f_{\rm A}$ is counter-intuitive, as bridging chains constrain relaxation of slip surfaces (i.e., terminal boundaries).
We propose that while bridging chains indeed increase the stiffness of the melt for short-to-intermediate times, their softening effect over longer timescales comes from (i) reduction in the average stretch of chains under deformation due to depletion of the looping chains at the terminal boundary; (ii) long-time relaxation of the melt due to interconversion between bridging and looping chains and re-distribution of bridging chains.}
\edits{This can be rationalized by the depression of the $D$ spacing for ABA melts compared with AB melts at equivalent values of $\chi N$---a finite-segregation effect that vanishes as $\chi N \to \infty$.
Finally, we close the equilibrium characterization of the lamellar and columnar phases as versions of lyotropic liquid crystals by including estimates of bending moduli.
Our approach, which uses external fields to deflect domain shapes, differs from prior methods based on wrapping block copolymer bilayers into specific curved shapes~\cite{Wang1992,Feuz2005,Li2013,Zhang2015} and may be considered a finite-wavenumber linear response of the melt~\cite{Ranjan2008}.
Given the flexibility of the external field approach for probing melt properties, a more in-depth exploration of liquid crystal mechanics will be left for future work.
}


The equilibrium assumption of SCFT calculations requires that polymers have had sufficient time to sample a large number of microstates.
Thus, there is no real distinction between rubbery and glassy polymers.
However, the dynamical rigidity plays an important role in the design of widely-used BCPs such as PDMS-PEO (rubbery-rubbery), SBS (glassy-rubbery), and PS-PMMA (glassy-glassy).
To experimentally corroborate our results, ideally rubbery BCPs (amorphous polymers with glass transition temperatures below room temperature --- i.e., PB, PEO, PDMS, amorphous PE) may be synthesized with targeted microphase based on $\chi N$, $f_{\rm A}$ and Kuhn length ratio $\epsilon$ and characterized via TEM or SAXS, with deformations studied under shear rheology at various temperatures and shear rates. 
Time-temperature superposition can provide $\chi$ (if unknown) for higher frequencies and also provide low-frequency modulus, which can be compared with the results discussed herein and possibly build out the full frequency-modulus plot as illustrated in Figure~\ref{fig:scheme_def_overview}.
A detailed description of helpful methodologies to characterize the effect of polymer microstructure on deformation response is outlined by Honeker and Thomas~\cite{Honeker1996Impact} and would be useful to further advance this research.
A more detailed understanding of BCP \edits{equilibrium} mechanics with respect to microstructure and associated design parameters (segregation strength, volume fraction and Kuhn length of respective blocks) is advantageous for polymer systems for use in extended lifetime or high temperature applications, such as structural engineering designs (buildings, bridges) or automotive industry (break and engine parts).
This study paves the way to understanding failure mechanics in BCP melts, addressing creep in shear and extension modes.
Using the inherent microphase separation of BCP melts to achieve arrested creep and enhanced toughness has several advantages over chemically-crosslinked thermoset materials, including no need for potentially toxic additives (e.g.,~crosslinkers and initiators) and enhanced degradability profile.

Finally, this work can be regarded as a study of mass transport and how equilibrium mass exchange between domains and along interfaces---known to be highly morphology-dependent~\cite{Shen2018}---determines the quasistatic response of polymer melts.
The liquid-like response of lamellar and columnar morphologies, along with long-time activated exchange of bridging chains between domains, helps expose a picture of equilibration of BCP melts across scales that has been recently developed by theoretical~\cite{Reddy2018,Dimitriyev2024} and experimental~\cite{Feng2019,Feng2021,White2025} efforts.
We leave a more in-depth exploration of how domain structure, topology, and deformation affect chain exchange and loop-to-bridge interconversion for future study.

\begin{acknowledgement}

The authors would like to thank E.L.~Thomas, F.~Bates, G.M.~Grason, B.R.~Greenvall, and R.~Mathew for useful discussions. \edits{K.G.S.~was supported by the National Science Foundation Graduate Research Fellowship Program (NSF GRFP); G.R.~and M.S.D.~were supported in part by the National Science Foundation DMREF Progam (Grant No.~DMR-2522693).}

\end{acknowledgement}

\begin{suppinfo}

Supplementary text can be found at \msd{to be added}.
Data and code can be found at \msd{to be added}.

\end{suppinfo}

\bibliography{refs}

\end{document}



\title{Supplementary Information: \\Structural Relaxation and Anisotropic Elasticity of Ordered Block Copolymer Melts}

\author{Krista G.~Schoonover}
\affiliation{%
 Department of Chemistry, Texas A\&M University, College Station, TX 77843, USA
}

\author{Gaurav Rawat}
\affiliation{%
 Department of Materials Science and Engineering, Texas A\&M University, College Station, TX 77843, USA
}

\author{Emily B.~Pentzer}
\affiliation{%
 Department of Chemistry, Texas A\&M University, College Station, TX 77843, USA
}
\affiliation{%
 Department of Materials Science and Engineering, Texas A\&M University, College Station, TX 77843, USA
}

\author{Michael S.~Dimitriyev}
 \email{msdim@tamu.edu}
\affiliation{%
 Department of Materials Science and Engineering, Texas A\&M University, College Station, TX 77843, USA
}

\renewcommand\thefigure{S\arabic{figure}}
\renewcommand\thesection{S\arabic{section}}
\renewcommand\theequation{S\arabic{equation}}
\renewcommand\thetable{S\arabic{table}}

\maketitle


\begin{table}[ht]
\centering
    \begin{tabular}{l l c c c c}
    \hline
    Polymer & Structure & $C_\infty$ & $b$ (Å) & $\rho$ $(\rm{g cm}^{-3})$ & $M_0$ $(\rm{g mol}^{-1})$ \\
    \hline
    1,4-Polyisoprene (PI) & --(CH$_2$CH=CHCH(CH$_3$))-- & 4.6 & 8.2 & 0.830 & 113 \\
    1,4-Polybutadiene (PB) & --(CH$_2$CH=CHCH$_2$)-- & 5.3 & 9.6 & 0.826 & 105 \\
    Polypropylene (PP) & --(CH$_2$CH$_2$(CH$_3$))-- & 5.9 & 11 & 0.791 & 180 \\
    Poly(ethylene oxide) (PEO) & --(CH$_2$CH$_2$O)-- & 6.7 & 11 & 1.064 & 137 \\
    Poly(dimethyl siloxane) (PDMS) & --(OSi(CH$_3$)$_2$)-- & 6.8 & 13 & 0.895 & 381 \\
    Polyethylene (PE) & --(CH$_2$CH$_2$)-- & 7.4 & 14 & 0.784 & 150 \\
    Poly(methyl methacrylate) & --(CH$_2$C(CH$_3$)(COOCH$_3$))-- & 9.0 & 17 & 1.13 & 655 \\
    Atactic polystyrene (PS) & --(CH$_2$CHC$_6$H$_5$)-- & 9.5 & 18 & 0.969 & 720 \\
    \hline
    \end{tabular}
    \caption{Characteristic ratios ($C_\infty$), Kuhn lengths ($b$), densities ($\rho$), and molar masses of Kuhn monomers ($M_0$) for common polymers. Reproduced from Rubinstein textbook \cite{Rubinstein2003}.}
    \label{tab:kuhn_lengths}
\end{table}

\section{Parameters used for SCFT calculations}

In order to allow for a more general set of relaxation modes under deformation, we lifted constraints on the symmetry groups of the microphase separated patterns.
This is done after first determining the free energy, monomer distributions, and chemical potential fields in the highest symmetry, undeformed ``reference'' configurations.
For details about the symmetry group conversions, see table~\ref{tab:symmetry_group}.

\begin{table}
  \centering
  \begin{tabular}{|c | c | c | c | c|}
    \hline
    & \multicolumn{2}{c|}{Undeformed (``reference'') configuration} & \multicolumn{2}{c|}{Deformed (``target'') configuration} \\
    \hline
    \#D & Lattice Type & Symmetry Group & Lattice Type & Symmetry Group \\
    \hline
    1D & lamellar & $P\bar{1}$ & lamellar & $P\bar{1}$ \\
    2D & hexagonal & $p6mm$ & oblique & $p1$ \\
    3D & cubic & $Ia\bar{3}d$ // $Im\bar{3}m$ & triclinic & $P1$ \\
    \hline
  \end{tabular}
  \caption{Table showing the conversions between symmetry groups for the undeformed-deformed configurations.}
  \label{tab:symmetry_group}
\end{table}

For the internal parameters used by the \texttt{PSCF} software~\cite{arora_broadly_2016}, we used the following values:
\begin{itemize}
    \item The internal \texttt{ds} parameter (corresponding to step sizes in the integrals over the chain backbone) is given by 0.01, increased to 0.05 for most cubic deformations with negligible impact on results as tested on gyroid structure at $\epsilon =1$, $\chi N^{\rm ABA} = 30$. 
    \item Mesh size, specified by \texttt{mesh}, ranged from 40 for lamellae, $48 \times 48$ up to $80 \times 80$ for hexagonal cylinders, and $64 \times 64 \times 64$ up to $100 \times 100 \times 100$ for cubic structures.
    \item Maximum history, specified by \texttt{maxHist}, ranged from 10 (the set value during phase sweeps) to 30 for most deformations, with high $\chi N$ cubic structures requiring above 200 to converge.
    \item Convergence threshold, specified by \texttt{epsilon}, was set at $1\times 10^{-5}$ for all phase sweeps and deformations.
\end{itemize}

\vspace{5pt}
The deformation cell parameters are given by
\begin{equation}
    a = (1+s)a_0
\end{equation}
for lamellae, where $a_0$ is the initial lattice parameter and $s$ is the incremental deformation, here $s=0-0.1$ in increments of $0.01$. Hexagonal deformation cell parameters are given by 
\begin{equation}
\begin{split}
    a &= (1+s)a_0 \\
    b &= a_0 \\
    \gamma &= \frac{2\pi}{3}
\end{split}
\end{equation}
for uniaxial stretch and 
\begin{equation}
\begin{split}
    a &= b = (1+s)a_0 \\ 
    \gamma &= \frac{2\pi}{3} 
\end{split}
\end{equation}
for isotropic biaxial stretch. A number of shear stretches were performed to evaluate monomer density relaxation as follows:
\begin{equation}
\begin{split}
    a &= b = a_0 \\
    \gamma &= (1+s)\frac{2\pi}{3}.
\end{split}
\end{equation}
Deformations for cubic geometries are given by the following lattice parameters for simple shear:
\begin{equation}
\begin{split}
    a &= c = a_0 \\
    b &= \frac{a_0}{\sin{\gamma}} \\
    \alpha &= \beta = 0 \\
    \gamma &= \arctan(\frac{1}{s}).
\end{split}
\end{equation}
and the following lattice parameters for simple extension:
\begin{equation}
\begin{split}
    a &= (1+s)a_0 \\
    b &= c = \frac{a_0}{\sqrt{1+s}} \\
    \alpha &= \beta = 0 \\
    \gamma &= \frac{\pi}{2}.
\end{split}
\end{equation}

\section{Extracting 2D moduli from SCFT}

Two protocols are used here: uniaxial stretching of one axis and isotropic biaxial stretching in 2D.

\subsection{Uniaxial stretch}

In the case of uniaxial stretching the lattice parameters are determined as a function of the stretch parameter $s$ as follows:
\begin{equation}
    \begin{split}
        a &= a_0(1 + s) \\
        b &= a_0 \\
        \theta &= \frac{2\pi}{3}
    \end{split}
\end{equation}
Therefore, the strain is given by
\begin{equation}
\begin{split}
    \bstrain \approx \left(\begin{array}{cc}
    s & 0 \\
    0 & 0
    \end{array}\right) 
\end{split}
\end{equation}
The deformation free energy per volume is then given by
\begin{equation}\begin{split} \label{uniaxial_energy}
    \Delta \mathcal{F} &= \frac{1}{2}\left[2\mu_{\perp} (\strain_{xx}^2 + \strain_{yy}^2 + 2\strain_{xy}\strain_{yx}) + \lambda_{\perp}(\strain_{xx} + \strain_{yy})^2\right] \\
    &= \frac{1}{2}\left[2\mu_{\perp} s^2 + \lambda_\perp s^2\right] \\
    &= \frac{1}{2}(2\mu_{\perp} + \lambda_\perp) s^2 = \frac{1}{2}A_1 s^2
\end{split}\end{equation}

\subsection{Isotropic biaxial stretch}

In the case of uniaxial stretching the lattice parameters are determined as a function of the stretch parameter $s$ as follows:
\begin{equation}
    \begin{split}
        a &= a_0(1 + s) \\
        b &= a_0(1 + s) \\
        \theta &= \frac{2\pi}{3}
    \end{split}
\end{equation}
Therefore, the strain is given by
\begin{equation}
    \bstrain \approx \left(\begin{array}{cc}
    s & 0 \\
    0 & s
    \end{array}\right)
\end{equation}
The deformation free energy per volume is then given by
\begin{equation}\begin{split} \label{isotropic_biaxial_energy}
    \Delta \mathcal{F} &= \frac{1}{2}\left[2\mu_{\perp} (\strain_{xx}^2 + \strain_{yy}^2 + 2\strain_{xy}\strain_{yx}) + \lambda_{\perp}(\strain_{xx} + \strain_{yy})^2\right] \\
    &= \frac{1}{2}\left[4\mu_{\perp} s^2 + 4\lambda_\perp s^2\right] \\
    &= 2(\mu_{\perp} + \lambda_\perp) s^2 = \frac{1}{2}A_2 s^2
\end{split}\end{equation}

\subsection{Deriving moduli from applied stretches}
Together, the applied uniaxial and isotropic biaxial stretches provide a system of 2 equations with 2 unknowns, for which we can solve $\lambda_\perp$ and $\mu_\perp$.
Subtracting $2\times$ \ref{uniaxial_energy} from \ref{isotropic_biaxial_energy} gives
\begin{equation}
    \lambda_\perp = -A_1+\frac{1}{2}A_2
\end{equation}
and subtracting \ref{isotropic_biaxial_energy} from $4\times$ \ref{uniaxial_energy} gives
\begin{equation}
    \mu_\perp = A_1-\frac{1}{4}A_2
\end{equation}
The shear modulus is given by $\mu$ and the Young's modulus by $Y_\parallel=\lambda_\perp+\mu_\perp$.
For any $f_{\rm A}$ with converged uniaxial and isotropic biaxial stretches, we can now calculate $Y_\parallel$ and $\mu_\perp$ for the 2D case.

\eject

\section{Rigidity figures and tables}

\begin{figure}[h!]
    \centering
    \includegraphics[width=0.8\linewidth]{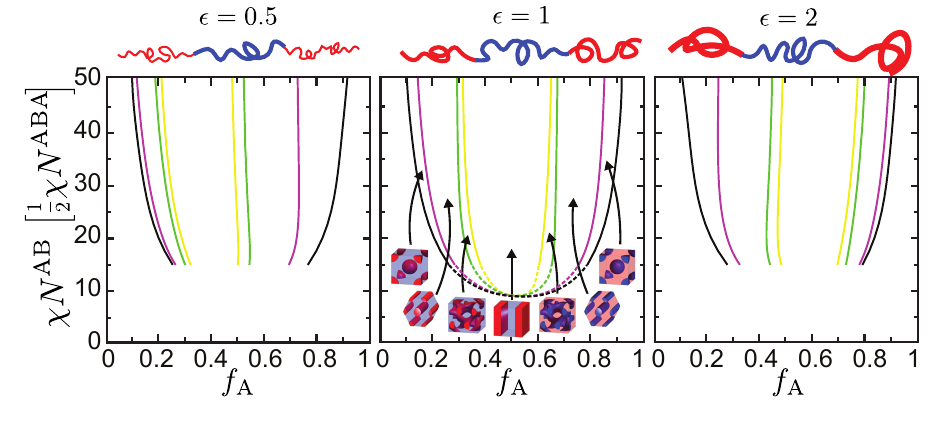}
    \caption{Phase diagrams for all $\epsilon = 0.5$, 1 and 2.}
    \label{fig:phase_diagram}
\end{figure}

\subsection{Lamellae}

\begin{figure}[h!]
    \centering
    \includegraphics[width=0.8\linewidth]{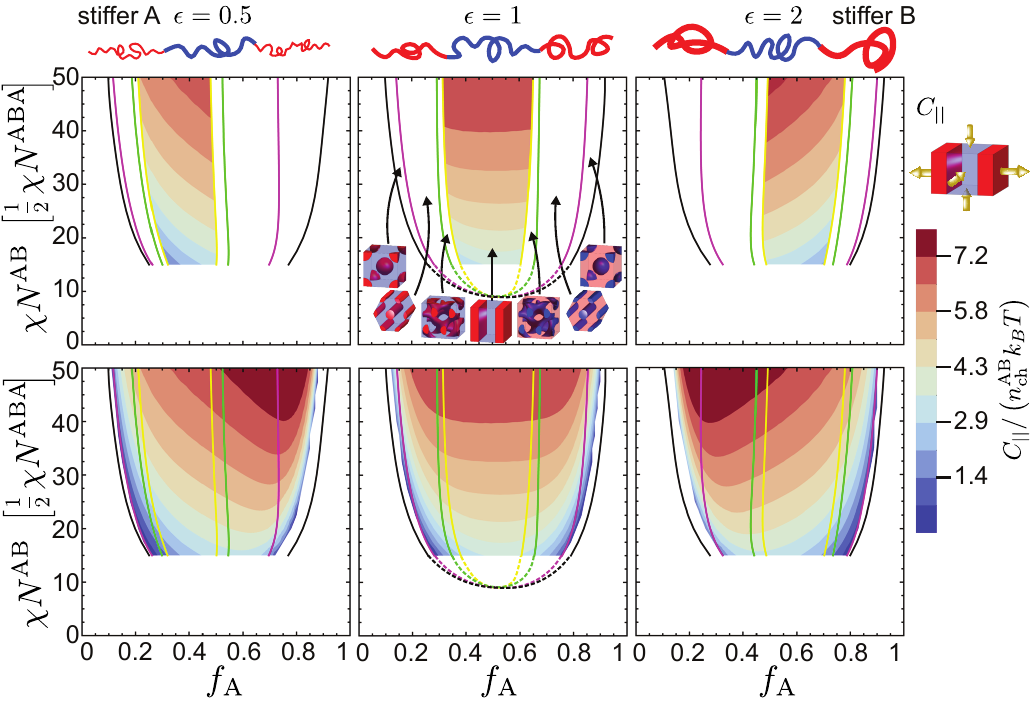}
    \caption{Lamellar heat maps of longitudinal modulus for ABA, comparing results within the stability window (top) to the full metastable region (bottom).}
    \label{fig:lam_mod_SI}
\end{figure}

\begin{figure}[h!]
    \centering
    \includegraphics[width=0.8\linewidth]{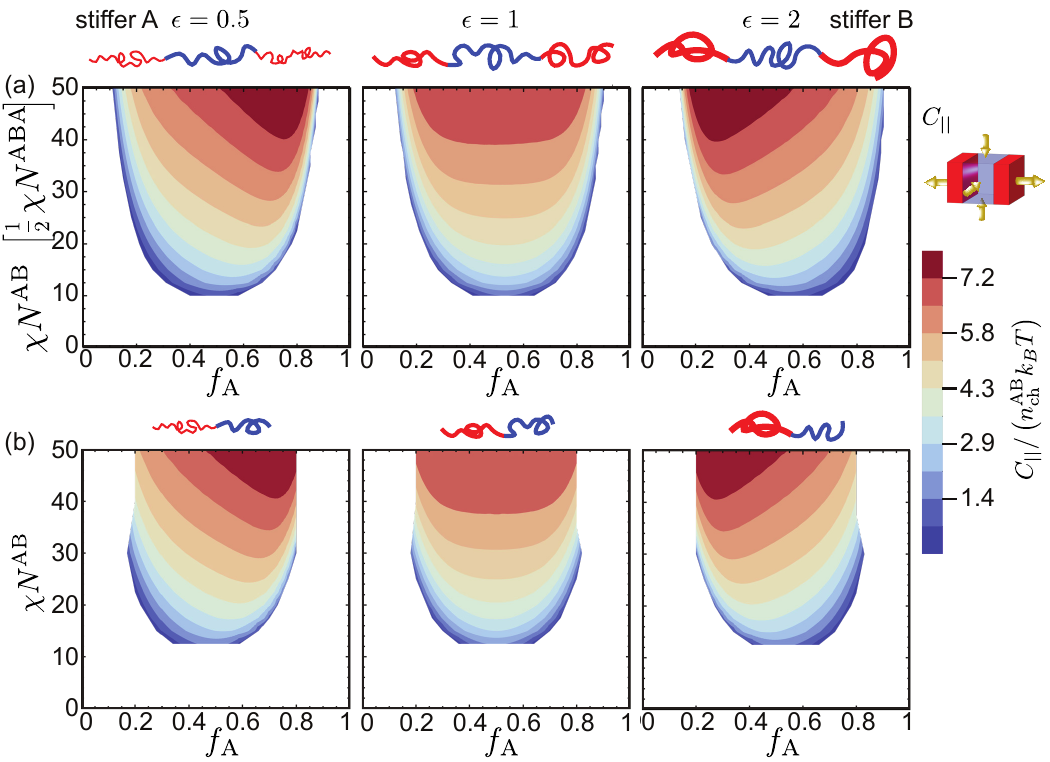}
    \caption{Lamellar heat maps of longitudinal modulus for (a) ABA and (b) AB, plotted on the same modulus scale.}
    \label{fig:lam_mod_full}
\end{figure}

\begin{figure}[h!]
    \centering
    \includegraphics[width=0.8\linewidth]{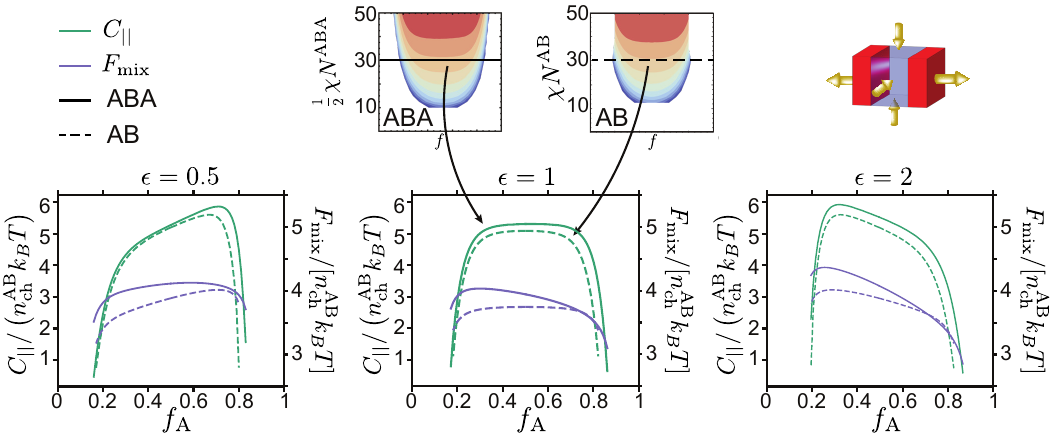}
    \caption{Lamellar plots of longitudinal modulus and free energy vs $f_{\rm A}$ for $\chi N^{\rm AB}=30$, comparing ABA and AB.}
    \label{fig:lam_mod_slice}
\end{figure}

\eject

\begin{figure}[h!]
    \centering
    \includegraphics[width=0.8\linewidth]{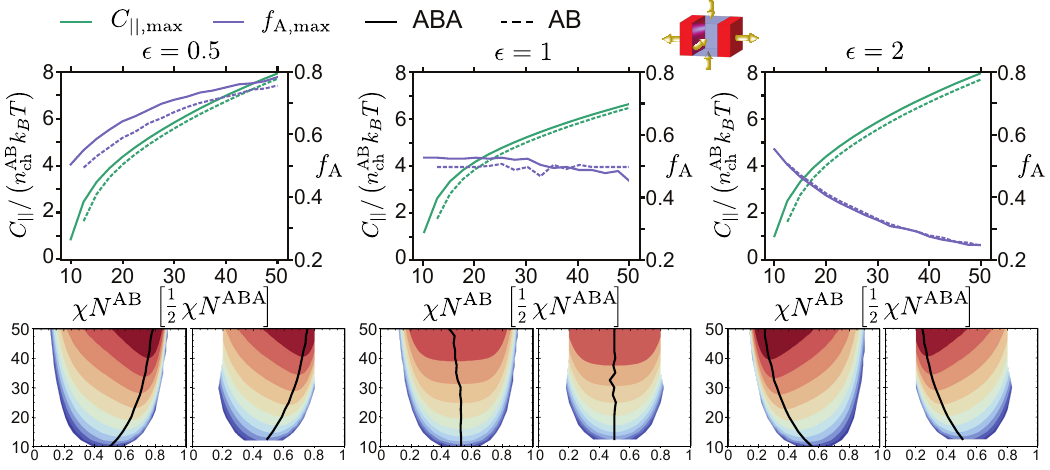}
    \caption{$C_{||,\rm{max}}$ and $f_{\rm{max}}$ for ABA and AB lamellae, plotted against $\chi N^{\rm AB}$.}
    \label{fig:lam_max_mod}
\end{figure}

The maximum longitudinal moduli, $C_{||,\rm{max}}$, and $f_{\rm A}$ at $C_{||,\rm{max}}$ are shown in Tables \ref{tab:lam_maxMod_full_ABA}
(for ABA) and \ref{tab:lam_maxMod_full_AB} (for AB).
It is our hope that these values can serve as a useful guide for future theoretical and experimental studies.
A graphical representation of $C_{||,\rm{max}}$ and associated $f_{\rm{A}}$ is shown in Figure \ref{fig:lam_max_mod}.
Notably, $f_{\rm{A}}$ have some numerical fluctuations at high $\chi N$ for ABA and AB for $\epsilon=1$, but in general $f_{\rm{A}}^{\rm AB}=0.5$ whereas $f_{\rm{A}}^{\rm ABA}\neq0.5$.
For configurational asymmetry, $f_{\rm{A}}$ increases with $\chi N$ for $\epsilon=0.5$ with a greater effect for ABA than AB, whereas $f_{\rm{A}}$ decreases with $\chi N$ for $\epsilon=2$ and is overlapping for ABA and AB.

\begin{table}[h!]
\centering
\begin{tabular}{|c|c c|c c|c c|} 
\hline
ABA$_{1\rm D}$ &\multicolumn{2}{c|}{$\epsilon=0.5$} & \multicolumn{2}{c|}{$\epsilon=1$} & \multicolumn{2}{c|}{$\epsilon=2$} \\
$\chi N$ &$C_{||,\rm{max}}$ & $f_{\rm A}$ & $C_{||,\rm{max}}$ & $f_{\rm A}$ & $C_{||,\rm{max}}$ & $f_{\rm A}$ \\
\hline
12.5 & 2.439 & 0.549 & 2.655 & 0.530 & 2.499 & 0.504 \\
15 & 3.281 & 0.585 & 3.386 & 0.527 & 3.322 & 0.465 \\
17.5 & 3.876 & 0.615 & 3.868 & 0.527 & 3.919 & 0.435 \\
20 & 4.359 & 0.642 & 4.238 & 0.530 & 4.408 & 0.405 \\
22.5 & 4.777 & 0.658 & 4.546 & 0.527 & 4.831 & 0.382 \\
25 & 5.158 & 0.678 & 4.818 & 0.530 & 5.215 & 0.362 \\
27.5 & 5.508 & 0.698 & 5.062 & 0.523 & 5.566 & 0.342 \\
30 & 5.835 & 0.711 & 5.289 & 0.527 & 5.895 & 0.325 \\
32.5 & 6.147 & 0.721 & 5.499 & 0.507 & 6.204 & 0.306 \\
35 & 6.444 & 0.734 & 5.700 & 0.497 & 6.499 & 0.299 \\
37.5 & 6.721 & 0.741 & 5.889 & 0.500 & 6.776 & 0.289 \\
40 & 6.992 & 0.751 & 6.071 & 0.490 & 7.037 & 0.272 \\
42.5 & 7.242 & 0.761 & 6.243 & 0.490 & 7.293 & 0.262 \\
45 & 7.489 & 0.764 & 6.405 & 0.480 & 7.538 & 0.256 \\
47.5 & 7.722 & 0.771 & 6.564 & 0.487 & 7.767 & 0.246 \\
50 & 7.947 & 0.784 & 6.716 & 0.450 & 7.992 & 0.246 \\
\hline
\end{tabular}
\caption{$f_{\rm A}$ and modulus in units of $n_{\rm{ch}}^{\rm AB}k_BT$ for the maximum longitudinal modulus for ABA lamellae.}
\label{tab:lam_maxMod_full_ABA}
\end{table}

\begin{table}[h!]
\centering
\begin{tabular}{|c|c c|c c|c c|} 
\hline
AB$_{1\rm D}$&\multicolumn{2}{c|}{$\epsilon=0.5$} & \multicolumn{2}{c|}{$\epsilon=1$} & \multicolumn{2}{c|}{$\epsilon=2$} \\
$\chi N$ &$C_{||,\rm{max}}$ & $f_{\rm A}$ & $C_{||,\rm{max}}$ & $f_{\rm A}$ & $C_{||,\rm{max}}$ & $f_{\rm A}$ \\
\hline
12.5 & 1.602 & 0.493 & 1.780 & 0.500 & 1.602 & 0.507 \\
15 & 2.758 & 0.529 & 2.852 & 0.500 & 2.758 & 0.471 \\
17.5 & 3.481 & 0.559 & 3.481 & 0.500 & 3.481 & 0.441 \\
20 & 4.027 & 0.589 & 3.923 & 0.500 & 4.027 & 0.411 \\
22.5 & 4.479 & 0.608 & 4.273 & 0.503 & 4.478 & 0.388 \\
25 & 4.879 & 0.635 & 4.570 & 0.510 & 4.880 & 0.368 \\
27.5 & 5.241 & 0.651 & 4.833 & 0.490 & 5.242 & 0.349 \\
30 & 5.580 & 0.671 & 5.073 & 0.500 & 5.579 & 0.329 \\
32.5 & 5.898 & 0.688 & 5.293 & 0.470 & 5.897 & 0.312 \\
35 & 6.197 & 0.701 & 5.502 & 0.507 & 6.197 & 0.299 \\
37.5 & 6.475 & 0.711 & 5.700 & 0.493 & 6.475 & 0.289 \\
40 & 6.745 & 0.721 & 5.883 & 0.507 & 6.745 & 0.276 \\
42.5 & 7.005 & 0.731 & 6.064 & 0.500 & 7.003 & 0.269 \\
45 & 7.251 & 0.744 & 6.233 & 0.500 & 7.249 & 0.256 \\
47.5 & 7.492 & 0.747 & 6.399 & 0.500 & 7.489 & 0.253 \\
50 & 7.721 & 0.757 & 6.553 & 0.500 & 7.720 & 0.243 \\
\hline
\end{tabular}
\caption{$f_{\rm A}$ and modulus in units of $n_{\rm{ch}}^{\rm AB}k_BT$ for the maximum longitudinal modulus for AB lamellae.}
\label{tab:lam_maxMod_full_AB}
\end{table}

\begin{figure}[h!]
    \centering
    \includegraphics[width=0.8\linewidth]{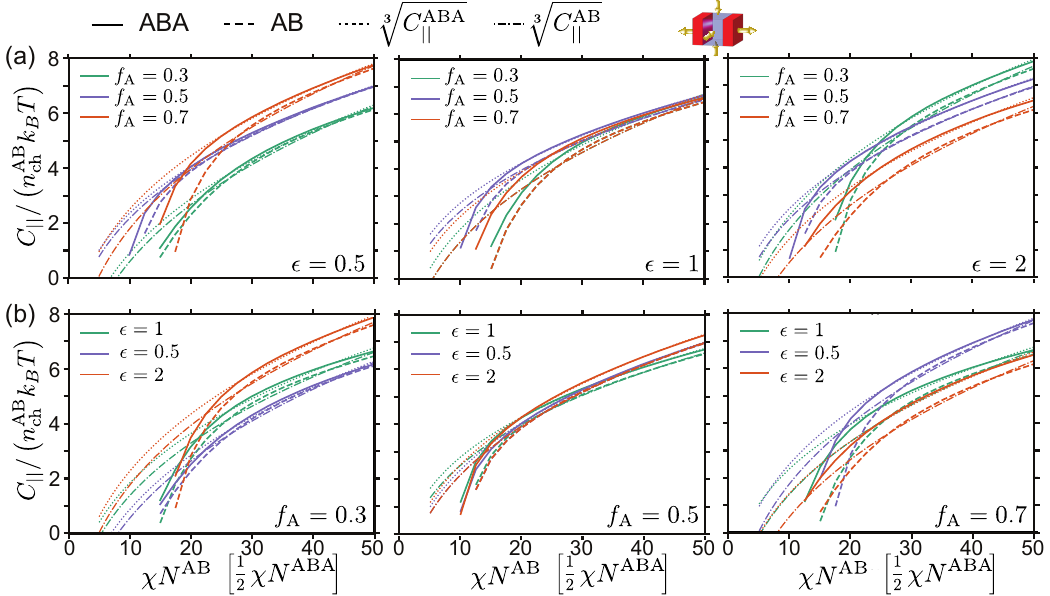}
    \caption{1D $\chi N$ vs longitudinal modulus, grouped by (a) $\epsilon$ or (b) $f_{\rm A}$ value for ABA and AB.}
    \label{fig:chiNvsY_1D}
\end{figure}

\clearpage

\subsection{Columnar Phase}

\begin{figure}[h!]
    \centering
    \includegraphics[width=0.8\linewidth]{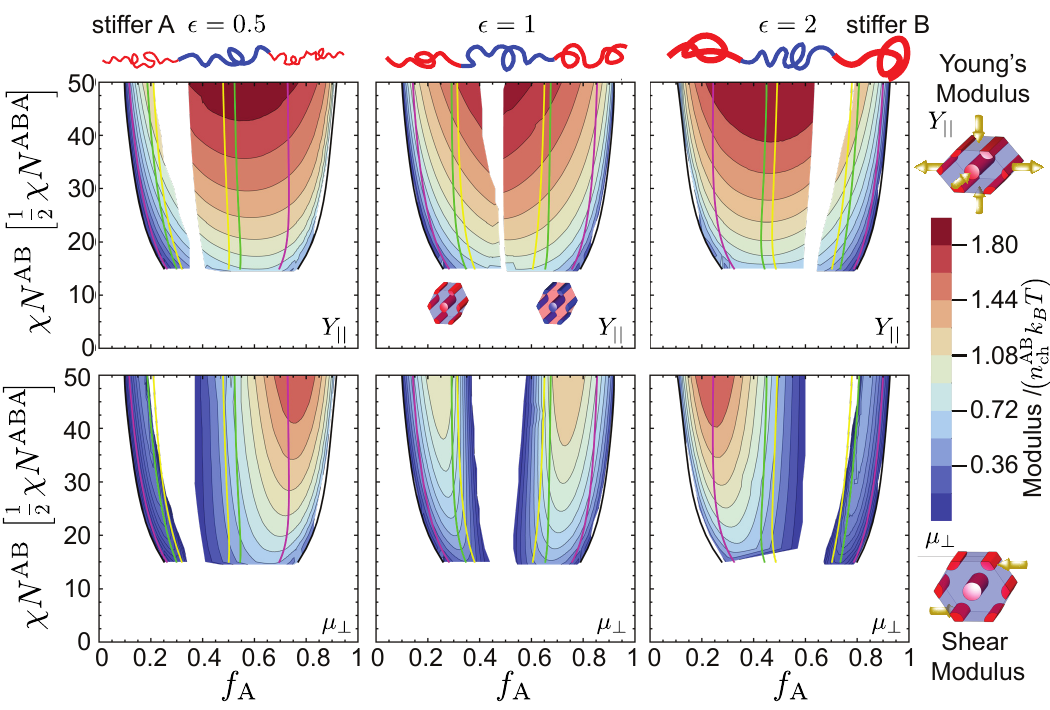}
    \caption{Full modulus heat map for ABA columnar phase, including metastable regions. Results plotted into middle of lamellar region, or to ends of converged structures.}
    \label{fig:hex_mod_full}
\end{figure}

\begin{figure}[h!]
    \centering
    \includegraphics[width=0.8\linewidth]{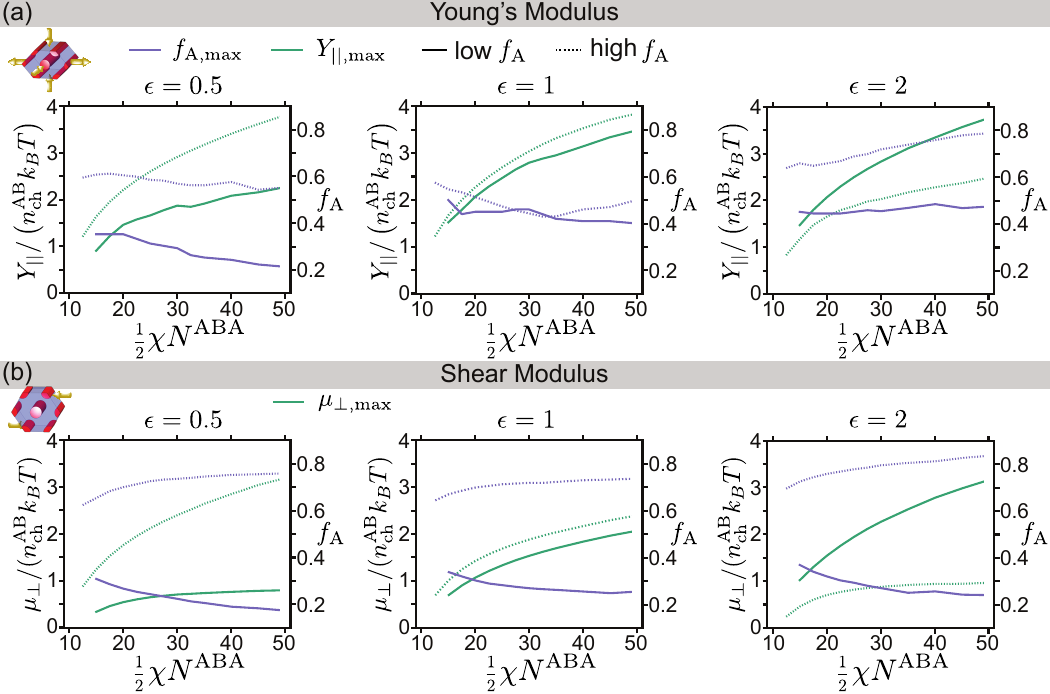}
    \caption{$Y_{||,\rm{max}}$ and $\mu_{\perp,\rm{max}}$ and their corresponding $f_{\rm{max}}$ for ABA cylinders, plotted against $\chi N$.}
    \label{fig:hex_max_mod}
\end{figure}

The maximum Young's and shear moduli, $Y_{||}$ and $\mu_\perp$, and $f_{\rm A}$ at the respective maxima are shown in Tables \ref{tab:hex1_maxMod}
- \ref{tab:hex3_maxMod}, grouped by $\epsilon$, with an overview for $\chi N^{\rm AB}=30$ shown in Table \ref{tab:hex_maxMod}.
A graphical representation of $Y_{||,\rm{max}}$, $f_{Y,\rm{max}}$ and $\mu_{\perp,\rm{max}}$, $f_{\mu,\rm{max}}$ is shown in Figure \ref{fig:hex_max_mod}.

\begin{table}[h!]
    \centering
    \begin{tabular}{|c||c c|c c||c c|c c|}
        \hline
        2D&\multicolumn{4}{c||}{ABA} & \multicolumn{4}{c|}{AB} \\
        \hline
        &\multicolumn{2}{c|}{low $f_{\rm A}$} & \multicolumn{2}{c||}{high $f_{\rm A}$} & \multicolumn{2}{c|}{low $f_{\rm A}$} & \multicolumn{2}{c|}{high $f_{\rm A}$} \\
        \hline
        $\epsilon$ &$\mu_{\perp}$ & $f_{\rm A}$ (bounds) & $\mu_{\perp}$ & $f_{\rm A}$ (bounds) & $\mu_{\perp}$ & $f_{\rm A}$ & $\mu_{\perp}$ & $f_{\rm A}$ \\
        \hline
        0.5 &0.35& 0.22 (0.16-0.22) & 1.21& 0.74 (0.53-0.73) & 0.41 & 0.23 & 1.08 & 0.72 \\
        1.0 &0.78& 0.27 (0.18-0.30) & 0.95& 0.72 (0.67-0.83) & 0.80 & 0.28 & 0.80 & 0.72 \\
        2.0 &1.14& 0.27 (0.25-0.44) & 0.44& 0.80 (0.78-0.86) & 1.08 & 0.28 & 0.41 & 0.77 \\
        \hline
    \end{tabular}
    \caption{Maximum shear modulus and associated $f_{\rm A}$ for the given $\epsilon$ at $\chi N^{\rm AB}=30$ for columnar ABA and AB.}
    \label{tab:hex_maxMod}
\end{table}

\begin{table}[h!]
\centering
\begin{tabular}{|c||c c|c c||c c|c c|} 
\hline
2D $\epsilon=1$&\multicolumn{4}{c||}{low $f_{\rm A}$ (B matrix)} & \multicolumn{4}{c|}{high $f_{\rm A}$ (A matrix)} \\
\hline
$\chi N^{\rm AB}$ &$Y_{||}$ & $f_{\rm A}$ & $\mu_\perp$ & $f_{\rm A}$ & $Y_{||}$ & $f_{\rm A}$ & $\mu_\perp$ & $f_{\rm A}$ \\
\hline
12.5&&  &  &  & 1.223 & 0.575 & 0.706 & 0.647 \\
\rowcolor{lightgray}[0.35cm][0.35cm]
15&1.524 & 0.500* & 0.711 & 0.340 & 1.675 & 0.546 & 1.029 & 0.673 \\
17.5&1.791 & 0.440* & 0.920 & 0.321 & 2.005 & 0.533 & 1.265 & 0.689 \\
\rowcolor{lightgray}[0.35cm][0.35cm]
20&2.072 & 0.450* & 1.087 & 0.304 & 2.274 & 0.513 & 1.445 & 0.702 \\
22.5&2.288 & 0.450* & 1.228 & 0.291 & 2.501 & 0.493 & 1.588 & 0.709 \\
\rowcolor{lightgray}[0.35cm][0.35cm]
25&2.471 & 0.450* & 1.350 & 0.284 & 2.700 & 0.473 & 1.706 & 0.716 \\
27.5&2.654 & 0.460* & 1.458 & 0.277 & 2.879 & 0.456 & 1.810 & 0.719 \\
\rowcolor{lightgray}[0.35cm][0.35cm]
30&2.799 & 0.460* & 1.552 & 0.271 & 3.041 & 0.443 & 1.901 & 0.722 \\
32.5&2.886 & 0.440* & 1.637 & 0.267 & 3.190 & 0.430* & 1.982 & 0.722 \\
\rowcolor{lightgray}[0.35cm][0.35cm]
35&2.957 & 0.420* & 1.715 & 0.264 & 3.323 & 0.430* & 2.055 & 0.726 \\
40&3.147 & 0.410* & 1.855 & 0.257 & 3.533 & 0.460* & 2.186 & 0.733 \\
\rowcolor{lightgray}[0.35cm][0.35cm]
45&3.342 & 0.410* & 1.981 & 0.250 & 3.719 & 0.470* & 2.312 & 0.736 \\
50&3.491 & 0.400* & 2.088 & 0.257 & 3.850 & 0.500* & 2.415 & 0.740 \\
\hline
\end{tabular}
\caption{Maximum shear and Young's moduli (in units of $n_{\rm{ch}}k_BT$) and their associated $f_{\rm A}$ for $\epsilon=1$ for ABA columnar. * corresponds to values that were not local maxima, but simply the limit of computational testing (deformations failed for B matrix $f*$). For reference, $f_{\rm A}$ stability region bounds for $\chi N^{\rm AB}=30$ are 0.18-0.30 for low $f_{\rm A}$ and 0.67-0.83 for high $f_{\rm A}$.}
\label{tab:hex1_maxMod}
\end{table}

\begin{table}[h!]
\centering
\begin{tabular}{|c||c c|c c||c c|c c|} 
\hline
2D $\epsilon=0.5$&\multicolumn{4}{c||}{low $f_{\rm A}$ (B matrix)} & \multicolumn{4}{c|}{high $f_{\rm A}$ (A matrix)} \\
\hline
$\chi N^{\rm AB}$ &$Y_{||}$ & $f_{\rm A}$ & $\mu_\perp$ & $f_{\rm A}$ & $Y_{||}$ & $f_{\rm A}$ & $\mu_\perp$ & $f_{\rm A}$ \\
\hline
12.5&&  &  &  & 1.198 & 0.593 & 0.866 & 0.625 \\
\rowcolor{lightgray}[0.35cm][0.35cm]
15&0.895 & 0.350* & 0.320 & 0.306 & 1.627 & 0.606 & 1.238 & 0.655 \\
17.5&1.208 & 0.350* & 0.444 & 0.283 & 1.944 & 0.610 & 1.526 & 0.685 \\
\rowcolor{lightgray}[0.35cm][0.35cm]
20&1.451 & 0.350* & 0.529 & 0.264 & 2.198 & 0.603 & 1.761 & 0.702 \\
22.5&1.567 & 0.330* & 0.589 & 0.250 & 2.411 & 0.597 & 1.957 & 0.715 \\
\rowcolor{lightgray}[0.35cm][0.35cm]
25&1.653 & 0.310* & 0.635 & 0.240 & 2.596 & 0.583 & 2.125 & 0.728 \\
27.5&1.766 & 0.300* & 0.667 & 0.230 & 2.760 & 0.580 & 2.275 & 0.735 \\
\rowcolor{lightgray}[0.35cm][0.35cm]
30&1.863 & 0.290* & 0.694 & 0.220 & 2.908 & 0.567 & 2.408 & 0.738 \\
32.5&1.838 & 0.260* & 0.709 & 0.209 & 3.044 & 0.560 & 2.528 & 0.742 \\
\rowcolor{lightgray}[0.35cm][0.35cm]
35&1.907 & 0.250* & 0.725 & 0.202 & 3.171 & 0.560 & 2.655 & 0.748 \\
40&2.076 & 0.240* & 0.751 & 0.186 & 3.407 & 0.574 & 2.863 & 0.755 \\
\rowcolor{lightgray}[0.35cm][0.35cm]
45&2.150 & 0.220* & 0.773 & 0.179 & 3.616 & 0.541 & 3.060 & 0.758 \\
50&2.263 & 0.210* & 0.786 & 0.169 & 3.806 & 0.551 & 3.210 & 0.762 \\
\hline
\end{tabular}
\caption{Maximum shear and Young's moduli (in units of $n_{\rm{ch}}k_BT$) and their associated $f_{\rm A}$ for $\epsilon=0.5$ for ABA columnar. * corresponds to values that were not local maxima, but simply the limit of computational testing (deformations failed for B matrix $f*$.) For reference, $f_{\rm A}$ stability region bounds for $\chi N^{\rm AB}=30$ are 0.16-0.22 for low $f_{\rm A}$ and 0.53-0.73 for high $f_{\rm A}$.}
\label{tab:hex2_maxMod}

\end{table}

\begin{table}[h!]
\centering
\begin{tabular}{|c||c c|c c||c c|c c|} 
\hline
2D $\epsilon=2$&\multicolumn{4}{c||}{low $f_{\rm A}$ (B matrix)} & \multicolumn{4}{c|}{high $f_{\rm A}$ (A matrix)} \\
\hline
$\chi N^{\rm AB}$ &$Y_{||}$ & $f_{\rm A}$ & $\mu_\perp$ & $f_{\rm A}$ & $Y_{||}$ & $f_{\rm A}$ & $\mu_\perp$ & $f_{\rm A}$ \\
\hline
12.5&&  &  &  & 0.840 & 0.640* & 0.250 & 0.697 \\
\rowcolor{lightgray}[0.35cm][0.35cm]
15&1.481 & 0.453 & 1.029 & 0.371 & 1.182 & 0.660* & 0.466 & 0.727 \\
17.5&1.807 & 0.446 & 1.309 & 0.342 & 1.485 & 0.650* & 0.612 & 0.746 \\
\rowcolor{lightgray}[0.35cm][0.35cm]
20&2.076 & 0.446 & 1.552 & 0.322 & 1.661 & 0.660* & 0.712 & 0.760 \\
22.5&2.306 & 0.446 & 1.763 & 0.305 & 1.803 & 0.670* & 0.778 & 0.770 \\
\rowcolor{lightgray}[0.35cm][0.35cm]
25&2.506 & 0.453 & 1.954 & 0.295 & 1.886 & 0.690* & 0.827 & 0.780 \\
27.5&2.682 & 0.460 & 2.126 & 0.282 & 1.989 & 0.700* & 0.861 & 0.787 \\
\rowcolor{lightgray}[0.35cm][0.35cm]
30&2.840 & 0.456 & 2.278 & 0.272 & 2.037 & 0.720* & 0.888 & 0.797 \\
35&3.123 & 0.470 & 2.545 & 0.252 & 2.182 & 0.740* & 0.927 & 0.807 \\
\rowcolor{lightgray}[0.35cm][0.35cm]
40&3.355 & 0.486 & 2.792 & 0.258 & 2.291 & 0.760* & 0.950 & 0.814 \\
45&3.578 & 0.469 & 2.993 & 0.244 & 2.379 & 0.780* & 0.953 & 0.828 \\
\rowcolor{lightgray}[0.35cm][0.35cm]
50&3.775 & 0.476 & 3.174 & 0.242 & 2.498 & 0.790* & 0.974 & 0.838 \\
\hline
\end{tabular}
\caption{Maximum shear and Young's moduli (in units of $n_{\rm{ch}}k_BT$) and their associated $f_{\rm A}$ for $\epsilon=2$ for ABA columnar. * corresponds to values that were not local maxima, but simply the limit of computational testing (deformations failed for B matrix $f*$.) For reference, $f_{\rm A}$ stability region bounds for $\chi N^{\rm AB}=30$ are 0.25-0.44 for low $f_{\rm A}$ and 0.78-0.86 for high $f_{\rm A}$.}
\label{tab:hex3_maxMod}

\end{table}

\begin{figure}[h!]
    \centering
    \includegraphics[width=0.8\linewidth]{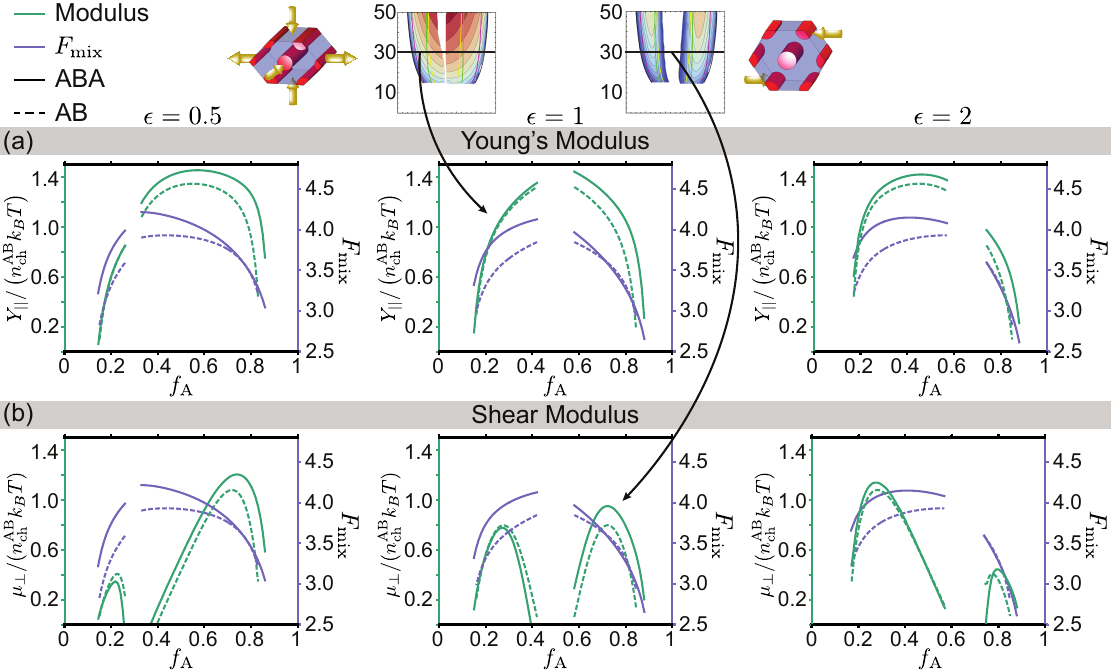}
    \caption{Plots of (a) Young's modulus and free energy and (b) shear modulus and free energy vs $f_{\rm A}$ for $\chi N^{\rm AB}=30$ for columnar phase, comparing ABA and AB.}
    \label{fig:hex_mod_slice}
\end{figure}

\begin{figure}[h!]
    \centering
    \includegraphics[width=0.8\linewidth]{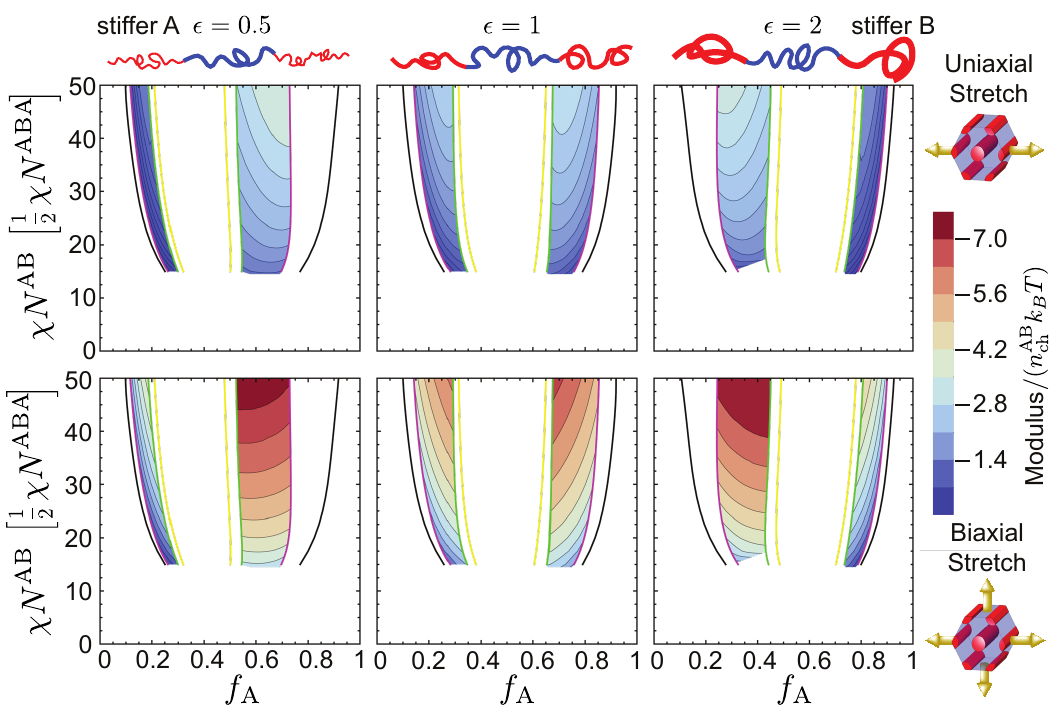}
    \caption{Modulus heat map for ABA columnar phase for original deformations (uniaxial and biaxial stretch), plotted within the stable phase regions.}
    \label{fig:hex_mod_orig}
\end{figure}

\begin{figure}[h!]
    \centering
    \includegraphics[width=0.8\linewidth]{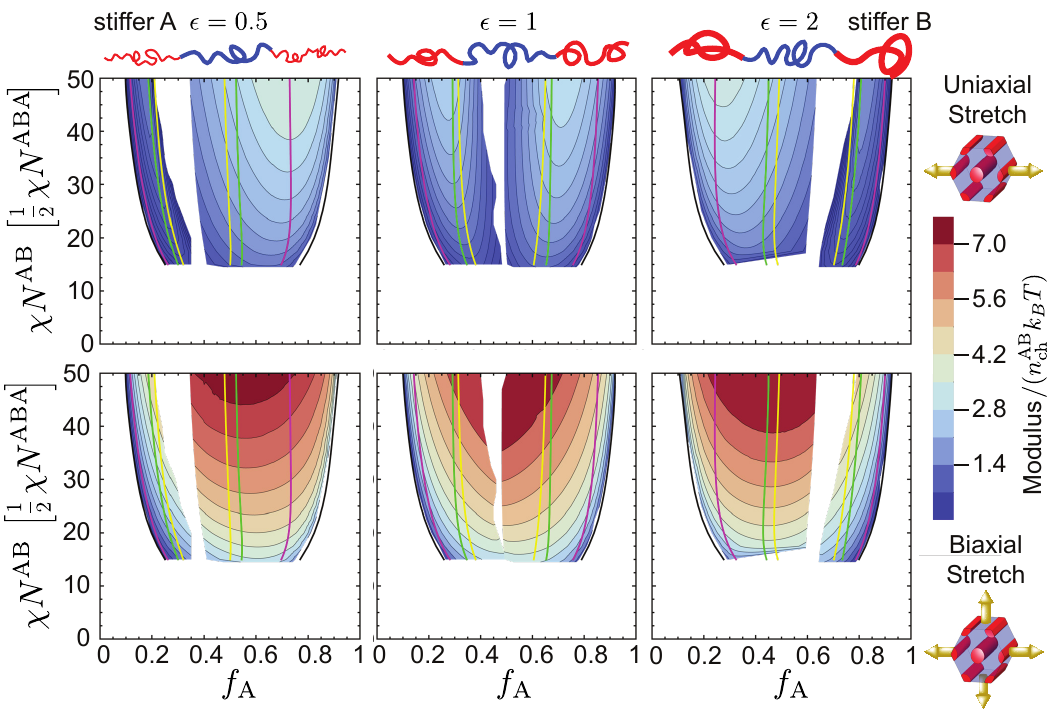}
    \caption{Full modulus heat map for ABA columnar phase for original deformations (uniaxial and biaxial stretch), including metastable regions. Results plotted into middle of lamellar region, or to ends of converged structures.}
    \label{fig:hex_mod_orig_full}
\end{figure}

\clearpage
\subsection{Cubic Phases}

\begin{figure}[h!]
    \centering
    \includegraphics[width=0.8\linewidth]{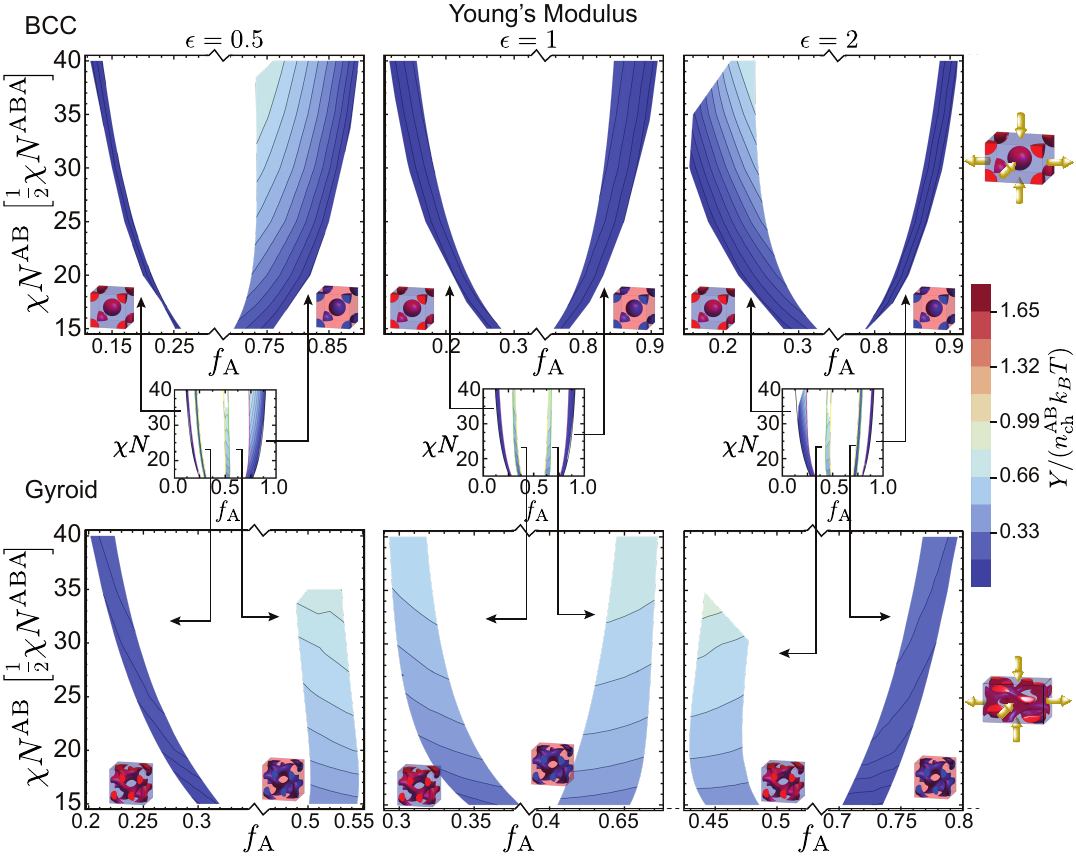}
    \caption{Modulus heat map for ABA 3D cubic phases for three elastic asymmetries for stable phase region, highlighting Young's modulus for BCC (top) and gyroid (bottom).}
    \label{fig:cubic_youngs_mod}
\end{figure}

\begin{figure}[h!]
    \centering
    \includegraphics[width=0.8\linewidth]{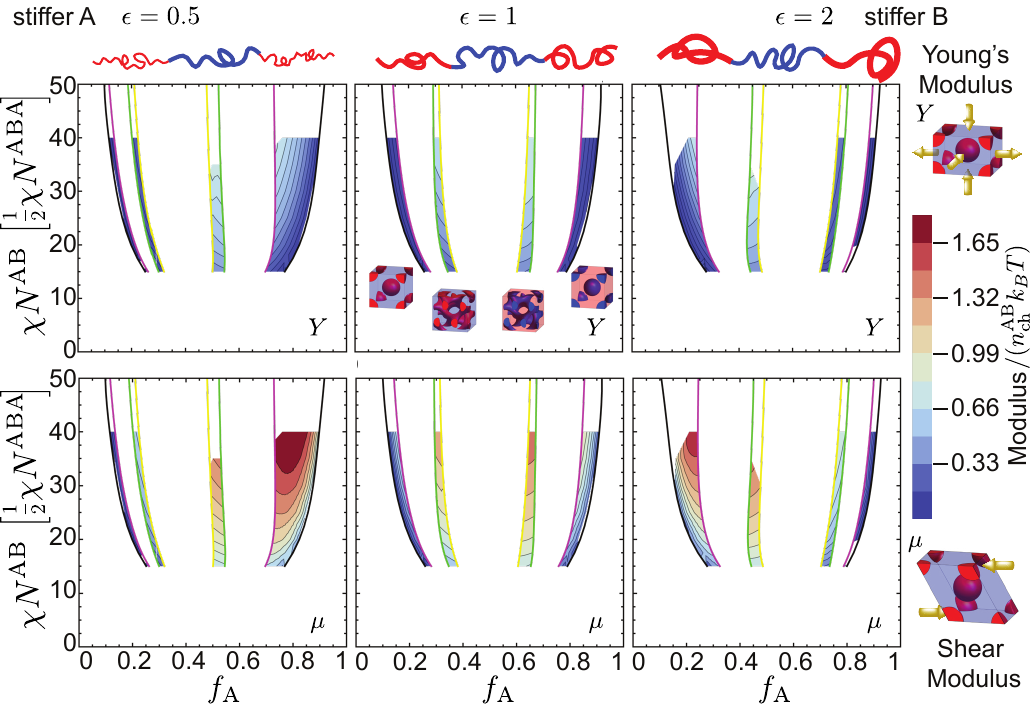}
    \caption{Modulus heat map for ABA 3D cubic phases for three elastic asymmetries for stable phase region, highlighting Young's modulus (top) and shear modulus (bottom) on regular scale $f_{\rm A}$.}
    \label{fig:cubic_mod}
\end{figure}

\begin{figure}[h!]
    \centering
    \includegraphics[width=0.8\linewidth]{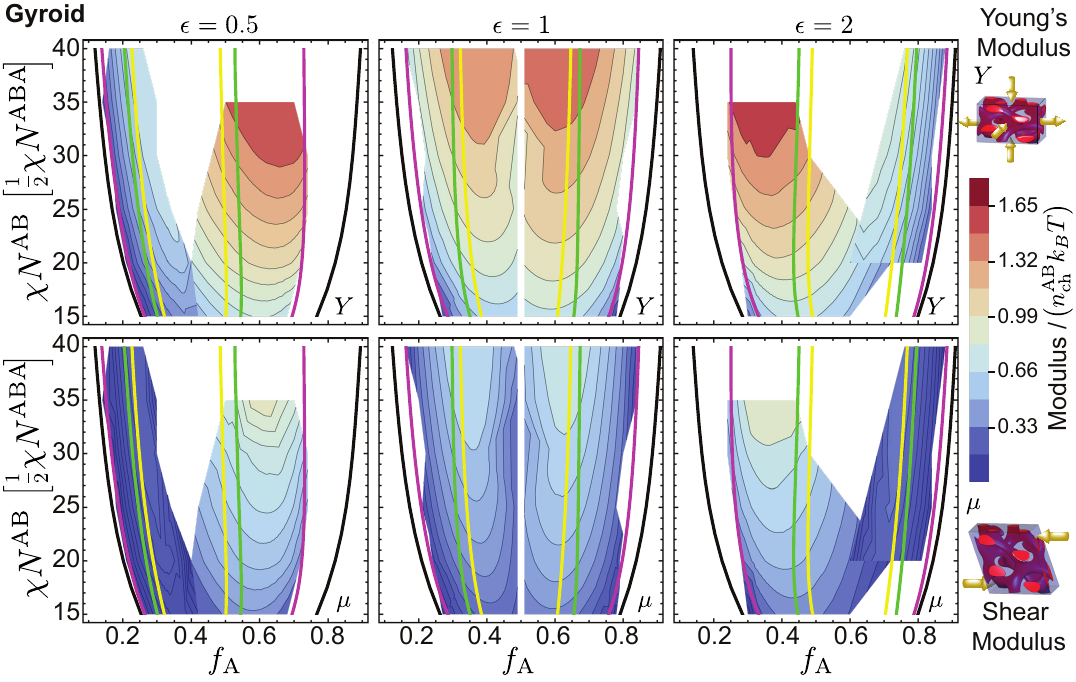}
    \caption{Modulus heat map for ABA gyroid phase for three elastic asymmetries including metastable regions, separated into Young's modulus (top) and shear modulus (bottom).}
    \label{fig:gyr_mod}
\end{figure}

\begin{figure}[h!]
    \centering
    \includegraphics[width=0.8\linewidth]{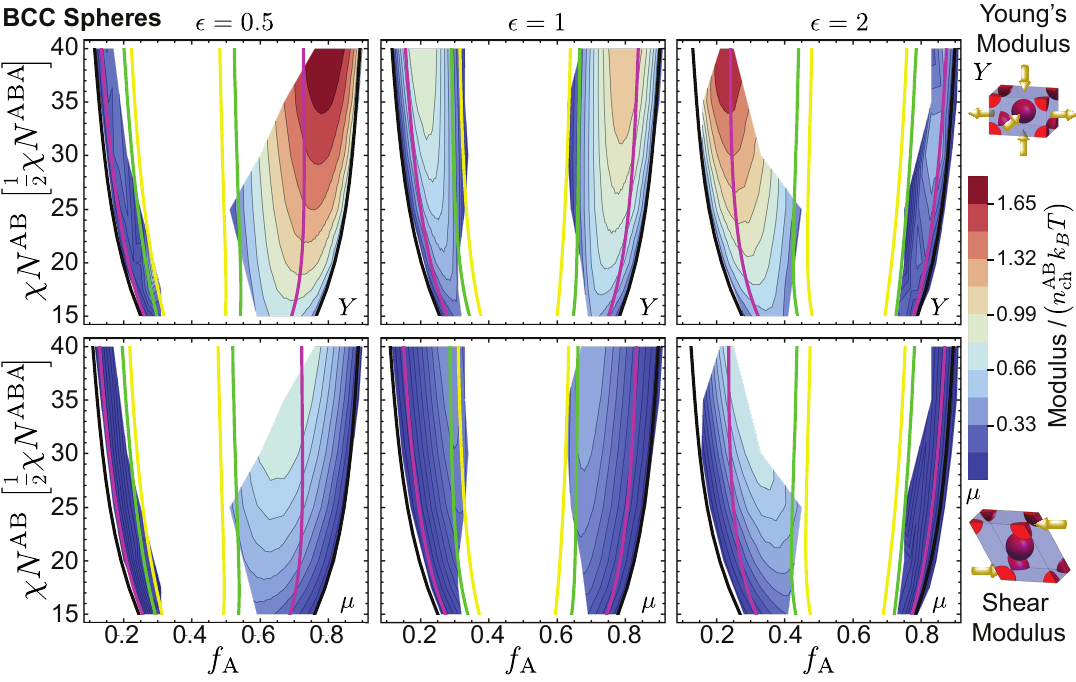}
    \caption{Modulus heat map for ABA BCC spheres for three elastic asymmetries including metastable regions, separated into Young's modulus (top) and shear modulus (bottom).}
    \label{fig:bcc_mod}
\end{figure}



\begin{figure}[h!]
    \centering
    \includegraphics[width=0.8\linewidth]{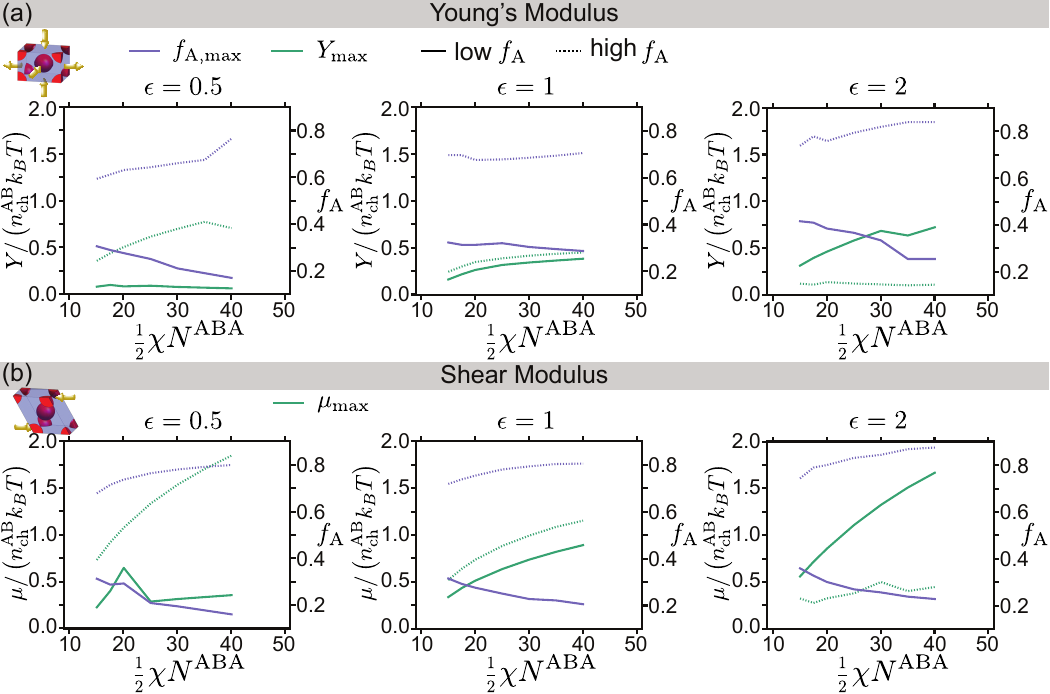}
    \caption{$Y_{\rm{max}}$ and $\mu_{\rm{max}}$ and their corresponding $f_{\rm{max}}$ for ABA BCC spheres, plotted against $\chi N$.}
    \label{fig:bcc_max_mod}
\end{figure}

\begin{table}[h!]
\centering
\begin{tabular}{|c||c c|c c||c c|c c|} 
\hline
BCC: $\epsilon=1$&\multicolumn{4}{c||}{low $f_{\rm A}$ (B matrix)} & \multicolumn{4}{c|}{high $f_{\rm A}$ (A matrix)} \\
\hline
$\chi N^{\rm AB}$ &$Y$ & $f_{\rm A}$ & $\mu$ & $f_{\rm A}$ & $Y$ & $f_{\rm A}$ & $\mu$ & $f_{\rm A}$ \\
\hline
15 & 0.305 & 0.410 & 0.549 & 0.354 & 0.114 & 0.730 & 0.316 & 0.735 \\
17.5 & 0.386 & 0.404 & 0.704 & 0.323 & 0.103 & 0.770 & 0.268 & 0.781 \\
20 & 0.453 & 0.379 & 0.844 & 0.296 & 0.130 & 0.750 & 0.317 & 0.792 \\
25 & 0.571 & 0.361 & 1.093 & 0.265 & 0.115 & 0.785 & 0.370 & 0.822 \\
30 & 0.673 & 0.329 & 1.307 & 0.252 & 0.106 & 0.810 & 0.490 & 0.835 \\
35 & 0.624 & 0.250 & 1.491 & 0.234 & 0.096 & 0.830 & 0.394 & 0.859 \\
40 & 0.711 & 0.250 & 1.648 & 0.224 & 0.103 & 0.830 & 0.437 & 0.866 \\
\hline
\end{tabular}
\caption{$f_{\rm A}$ and modulus in units of $n_{\rm{ch}}k_BT$ for the maximum shear and Young's modulus for $\epsilon=1$ for ABA BCC spheres. * corresponds to values that were not local maxima, but simply the limit of computational testing (deformations failed for B matrix $f*$).}
\label{tab:bcc1_maxMod}
\end{table}

\begin{table}
\centering
\begin{tabular}{|c||c c|c c||c c|c c|} 
\hline
BCC: $\epsilon=0.5$&\multicolumn{4}{c||}{low $f_{\rm A}$ (B matrix)} & \multicolumn{4}{c|}{high $f_{\rm A}$ (A matrix)} \\
\hline
$\chi N^{\rm AB}$ &$Y$ & $f_{\rm A}$ & $\mu$ & $f_{\rm A}$ & $Y$ & $f_{\rm A}$ & $\mu$ & $f_{\rm A}$ \\
\hline
15 & 0.356 & 0.590 & 0.724 & 0.672 & 0.082 & 0.304 & 0.222 & 0.310 \\
17.5 & 0.436 & 0.608 & 0.910 & 0.708 & 0.100 & 0.288 & 0.396 & 0.285 \\
20 & 0.502 & 0.627 & 1.066 & 0.730 & 0.085 & 0.275 & 0.640 & 0.290 \\
25 & 0.612 & 0.639 & 1.322 & 0.757 & 0.092 & 0.250 & 0.283 & 0.207 \\
30 & 0.695 & 0.656 & 1.522 & 0.773 & 0.079 & 0.210 & 0.310 & 0.193 \\
40 & 0.702 & 0.760 & 1.829 & 0.792 & 0.064 & 0.170 & 0.351 & 0.159 \\
\hline
\end{tabular}
\caption{$f_{\rm A}$ and modulus in units of $n_{\rm{ch}}k_BT$ for the maximum shear and Young's modulus for $\epsilon=0.5$ for ABA BCC spheres. * corresponds to values that were not local maxima, but simply the limit of computational testing (deformations failed for B matrix $f*$).}
\label{tab:bcc2_maxMod}
\end{table}

\begin{table}
\centering
\begin{tabular}{|c||c c|c c||c c|c c|} 
\hline
BCC: $\epsilon=2$&\multicolumn{4}{c||}{low $f_{\rm A}$ (B matrix)} & \multicolumn{4}{c|}{high $f_{\rm A}$ (A matrix)} \\
\hline
$\chi N^{\rm AB}$ &$Y$ & $f_{\rm A}$ & $\mu$ & $f_{\rm A}$ & $Y$ & $f_{\rm A}$ & $\mu$ & $f_{\rm A}$ \\
\hline
15 & 0.240 & 0.690 & 0.511 & 0.712 & 0.158 & 0.320 & 0.330 & 0.311 \\
17.5 & 0.295 & 0.690 & 0.630 & 0.731 & 0.213 & 0.310 & 0.426 & 0.288 \\
20 & 0.343 & 0.670 & 0.726 & 0.747 & 0.259 & 0.310 & 0.504 & 0.272 \\
25 & 0.382 & 0.672 & 0.872 & 0.773 & 0.312 & 0.317 & 0.627 & 0.247 \\
30 & 0.410 & 0.679 & 0.981 & 0.786 & 0.338 & 0.301 & 0.727 & 0.224 \\
35 & 0.430 & 0.688 & 1.074 & 0.796 & 0.359 & 0.292 & 0.810 & 0.218 \\
40 & 0.444 & 0.699 & 1.140 & 0.798 & 0.378 & 0.284 & 0.881 & 0.202 \\
\hline
\end{tabular}
\caption{$f_{\rm A}$ and modulus in units of $n_{\rm{ch}}k_BT$ for the maximum shear and Young's modulus for $\epsilon=2$ for ABA BCC spheres. * corresponds to values that were not local maxima, but simply the limit of computational testing (deformations failed for B matrix $f*$).}
\label{tab:bcc3_maxMod}
\end{table}



\begin{figure}[h!]
    \centering
    \includegraphics[width=0.8\linewidth]{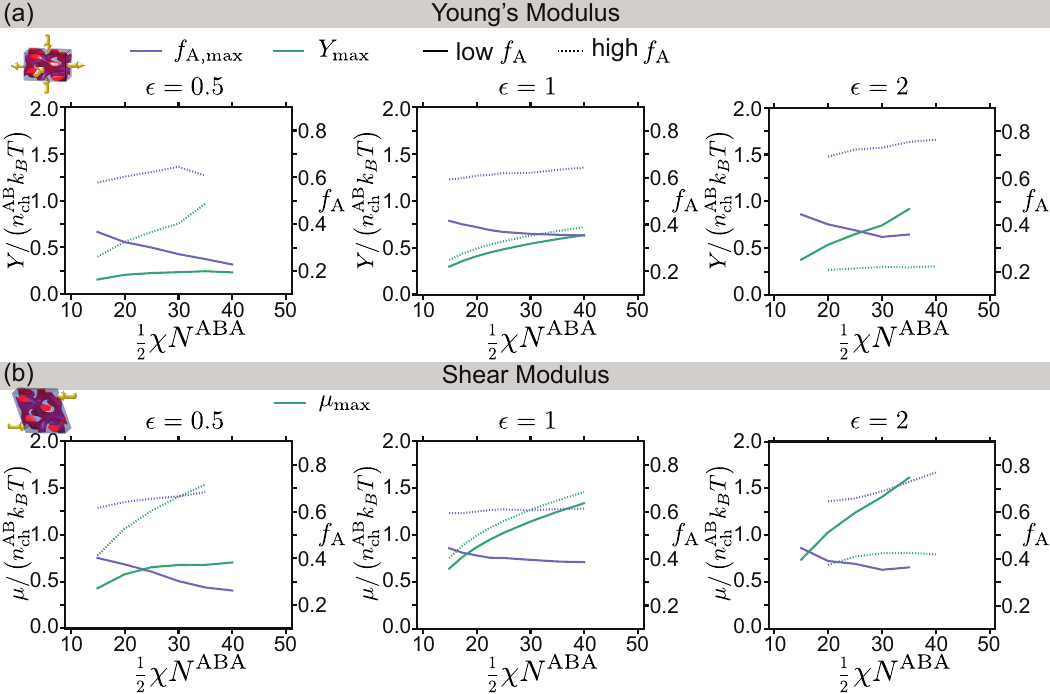}
    \caption{$Y_{\rm{max}}$ and $\mu_{\rm{max}}$ and their corresponding $f_{\rm{max}}$ for ABA gyroid, plotted against $\chi N$.}
    \label{fig:gyr_max_mod}
\end{figure}

\begin{table}[h!]
\centering
\begin{tabular}{|c||c c|c c||c c|c c|} 
\hline
GYR: $\epsilon=1$&\multicolumn{4}{c||}{low $f_{\rm A}$ (B matrix)} & \multicolumn{4}{c|}{high $f_{\rm A}$ (A matrix)} \\
\hline
$\chi N^{\rm AB}$ &$Y$ & $f_{\rm A}$ & $\mu$ & $f_{\rm A}$ & $Y$ & $f_{\rm A}$ & $\mu$ & $f_{\rm A}$ \\
\hline
15&0.610 & 0.416 & 1.266 & 0.439 & 0.752 & 0.591 & 1.489 & 0.588 \\
17.5 &0.730 & 0.401 & 1.515 & 0.420 & 0.888 & 0.597 & 1.756 & 0.588 \\
20 &0.827 & 0.390 & 1.712 & 0.409 & 0.991 & 0.607 & 1.958 & 0.594 \\
22.5 &0.908 & 0.377 & 1.876 & 0.399 & 1.074 & 0.611 & 2.124 & 0.601 \\
25 &0.977 & 0.371 & 2.018 & 0.397 & 1.146 & 0.618 & 2.267 & 0.605 \\
30 &1.094 & 0.357 & 2.257 & 0.380* & 1.267 & 0.619 & 2.508 & 0.600* \\
35 &1.191 & 0.357 & 2.466 & 0.383 & 1.362 & 0.632 & 2.709 & 0.605 \\
40 &1.278 & 0.354 & 2.647 & 0.380 & 1.448 & 0.642 & 2.885 & 0.607 \\
\hline
\end{tabular}
\caption{$f_{\rm A}$ and modulus in units of $n_{\rm{ch}}k_BT$ for the maximum shear and Young's modulus for $\epsilon=1$ for ABA gyroid. * corresponds to values that were not local maxima, but simply the limit of computational testing (deformations failed for B matrix $f*$).}
\label{tab:gyr1_maxMod}
\end{table}

\begin{table}[h!]
\centering
\begin{tabular}{|c||c c|c c||c c|c c|} 
\hline
GYR: $\epsilon=0.5$&\multicolumn{4}{c||}{low $f_{\rm A}$ (B matrix)} & \multicolumn{4}{c|}{high $f_{\rm A}$ (A matrix)} \\
\hline
$\chi N^{\rm AB}$ &$Y$ & $f_{\rm A}$ & $\mu$ & $f_{\rm A}$ & $Y$ & $f_{\rm A}$ & $\mu$ & $f_{\rm A}$ \\
\hline
15 &0.335 & 0.368 & 0.847 & 0.398 & 0.820 & 0.578 & 1.543 & 0.610 \\
20 &0.434 & 0.322 & 1.134 & 0.360 & 1.133 & 0.603 & 2.105 & 0.635 \\
25 &0.473 & 0.303 & 1.294 & 0.340 & 1.343 & 0.623 & 2.488 & 0.649 \\
30 &0.491 & 0.275 & 1.339 & 0.300 & 1.521 & 0.645 & 2.785 & 0.659 \\
35 &0.511 & 0.254 & 1.345 & 0.273 & 1.944 & 0.607 & 3.046 & 0.677 \\
40 &0.487 & 0.231 & 1.394 & 0.260 &  &  &  &  \\
\hline
\end{tabular}
\caption{$f_{\rm A}$ and modulus in units of $n_{\rm{ch}}k_BT$ for the maximum shear and Young's modulus for $\epsilon=0.5$ for ABA gyroid. * corresponds to values that were not local maxima, but simply the limit of computational testing (deformations failed for B matrix $f*$).}
\label{tab:gyr2_maxMod}
\end{table}

\begin{table}[h!]
\centering
\begin{tabular}{|c||c c|c c||c c|c c|} 
\hline
GRY: $\epsilon=2$&\multicolumn{4}{c||}{low $f_{\rm A}$ (B matrix)} & \multicolumn{4}{c|}{high $f_{\rm A}$ (A matrix)} \\
\hline
$\chi N^{\rm AB}$ &$Y$ & $f_{\rm A}$ & $\mu$ & $f_{\rm A}$ & $Y$ & $f_{\rm A}$ & $\mu$ & $f_{\rm A}$ \\
\hline
15 &0.759 & 0.443 & 1.454 & 0.440 &  &  &  &  \\
20 &1.073 & 0.401 & 2.034 & 0.385 & 0.541 & 0.688 & 1.341 & 0.638 \\
25 &1.299 & 0.376 & 2.448 & 0.373 & 0.573 & 0.718 & 1.519 & 0.650 \\
30 &1.484 & 0.348 & 2.784 & 0.348 & 0.603 & 0.726 & 1.590 & 0.680 \\
35 &1.829 & 0.358 & 3.191 & 0.358 & 0.594 & 0.750 & 1.600 & 0.720 \\
40 &&  &  &  & 0.616 & 0.760 & 1.564 & 0.760 \\
\hline
\end{tabular}
\caption{$f_{\rm A}$ and modulus in units of $n_{\rm{ch}}k_BT$ for the maximum shear and Young's modulus for $\epsilon=2$ for ABA gyroid. * corresponds to values that were not local maxima, but simply the limit of computational testing (deformations failed for B matrix $f*$).}
\label{tab:gyr3_maxMod}
\end{table}


\begin{figure}[h!]
    \centering
    \includegraphics[width=0.8\linewidth]{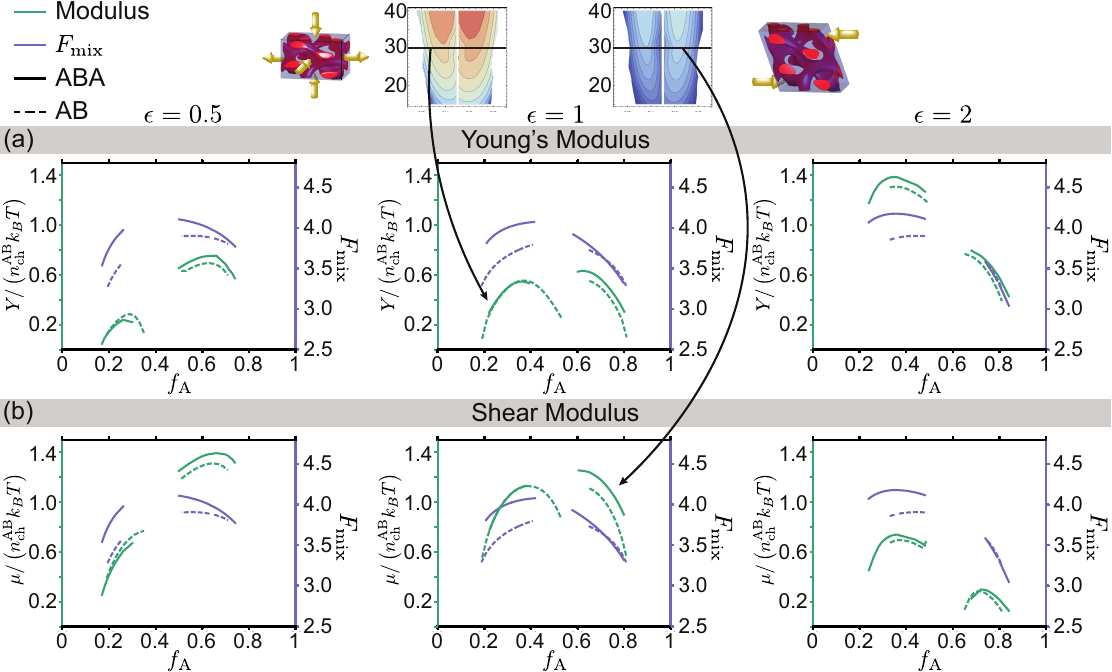}
    \caption{Plots of (a) Young's modulus and free energy and (b) shear modulus and free energy vs $f_{\rm A}$ for $\chi N^{\rm AB}=30$ for gyroid phase, comparing ABA and AB.}
    \label{fig:gyr_mod_slice}
\end{figure}

\begin{figure}[h!]
    \centering
    \includegraphics[width=0.8\linewidth]{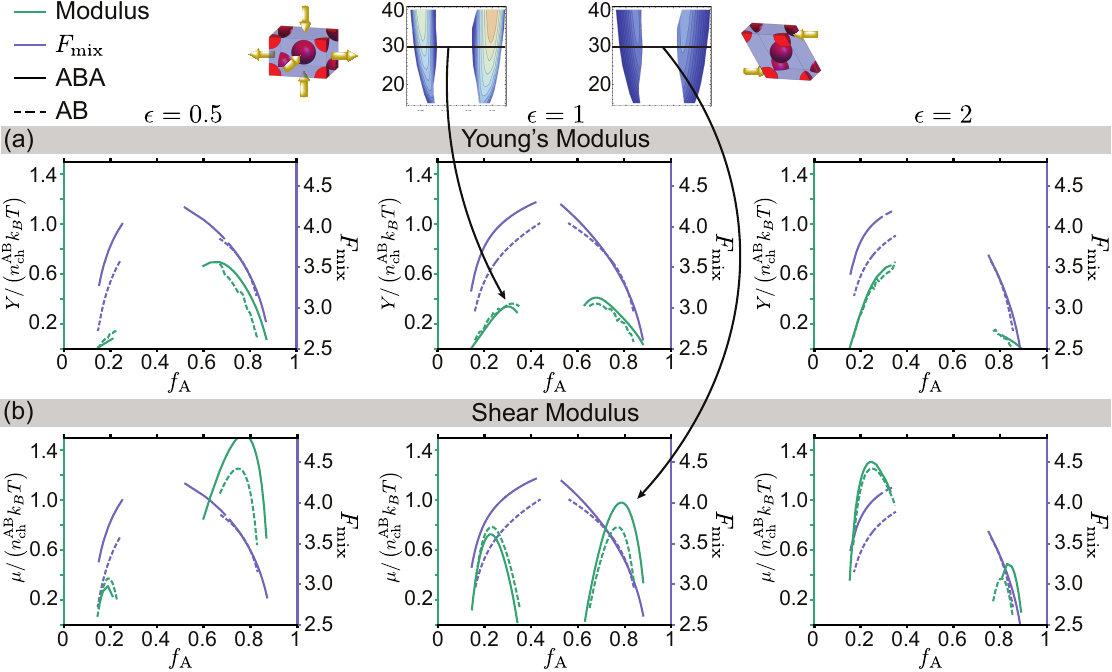}
    \caption{Plots of (a) Young's modulus and free energy and (b) shear modulus and free energy vs $f_{\rm A}$ for $\chi N^{\rm AB}=30$ for BCC spheres, comparing ABA and AB.}
    \label{fig:bcc_mod_slice}
\end{figure}


\clearpage


\section{Strong segregation theory calculations}

The strong segregation theory (SST) free energy is given by
\begin{equation*}
F^{\rm SST} = \gamma A + \frac{1}{2}\kappa_{\rm A}I_{\rm A} + \frac{1}{2}\kappa_{\rm B}I_{\rm B}
\end{equation*}
where $\gamma$ is an interfacial free energy density, $\kappa_{\rm A,B} = (3\pi^2 \rho k_B T)/(4N^2_{\rm A,B}\ell_{\rm A,B}^2)$ are entropic rigidities, and $I_{\rm A,B}$ are second moments of volume, corresponding to the geometric dependence of brush stretching from parabolic brush theory~\cite{Milner1988}.
For lamellae, the thickness of the A block brush is $h_{\rm A}$ and the thickness of the B block brush is $h_{\rm B}$.
Due to incompressibility of the melt, the brush thicknesses is related to the total lamellar thickness $h$ by $h_{\rm A} = f_{\rm A} h$ and $h_{\rm B} = (1-f_{\rm A}) h$.
The volume of the lamellar structure is given in terms of thickness $h = D/2$ and cross-section area $A$ by $V = AD$.
Then the second moments of volume are given by
\begin{align*}
I_{\rm A} &= \int{\rm d}V\, z^2 = A\int_0^{h_{\rm A}}{\rm d}z\, z^2 = \frac{1}{3}Ah_{\rm A}^3 = \frac{1}{24}Vf_{\rm A}^3D^2 \\ 
I_{\rm B} &= \int{\rm d}V\, z^2 = A\int_0^{h_{\rm B}}{\rm d}z\, z^2 = \frac{1}{3}Ah_{\rm B}^3 = \frac{1}{24}Vf_{\rm B}^3D^2
\end{align*}
The SST free energy of the unit cell is therefore
\begin{align*}
F^{\rm SST} &= 2\gamma A + \frac{1}{24}V\kappa_{\rm A}f_{\rm A}^3D^2 + \frac{1}{24}V\kappa_{\rm B}f_{\rm B}^3D^2 \\
&= \frac{2\gamma V}{D} + \frac{V}{24}\left[\frac{3\pi^2 \rho k_B T}{4 N_{\rm A}^2 \ell_{\rm A}^2}f^3_{\rm A} + \frac{3\pi^2 \rho k_B T}{4 N_{\rm B}^2 \ell_{\rm B}^2}f^3_{\rm B}\right]D^2 \\ 
&= \frac{2\gamma V}{D} + V\kappa\left[\frac{f_{\rm A}}{N\ell_{\rm A}^2} + \frac{f_{\rm B}}{N\ell_{\rm B}^2}\right]D^2
\end{align*}
where $\kappa \equiv \frac{\pi^2 \rho k_B T}{32 N} = (\pi^2/32)n_{\rm ch}k_B T$, where $n_{\rm ch} = \rho/N$ is the chain density.
The free energy per volume $\mathcal{F}^{\rm SST} = F^{\rm SST}/V$ is therefore
\begin{equation*}
\mathcal{F}^{\rm SST} = \frac{2\gamma}{D} + \kappa\left[\frac{f_{\rm A}}{N\ell_{\rm A}^2} + \frac{f_{\rm B}}{N\ell_{\rm B}^2}\right]D^2
\end{equation*}
In order to extract the elastic modulus at constant volume, consider the deformation free energy per chain $\Delta \mathcal{F}^{\rm SST}$ due to stretching the lamellae $D \mapsto \lambda D$, given by
\begin{align*}
\Delta \mathcal{F}^{\rm SST} &= \mathcal{F}^{\rm SST}(\lambda D) - \mathcal{F}^{\rm SST}(D) \\
&= \frac{2\gamma}{D}\left(\frac{1}{\lambda} - 1\right) + \kappa\left[\frac{f_{\rm A}}{N\ell_{\rm A}^2} + \frac{f_{\rm B}}{N\ell_{\rm B}^2}\right]D^2(\lambda^2 - 1)
\end{align*}
The elastic modulus $C_\parallel$ is therefore
\begin{align*}
C_\parallel^{\rm SST}(D) &= \frac{{\rm d}^2 \mathcal{F}^{\rm SST}}{{\rm d}\lambda^2}\bigg|_{\lambda = 1} \\ 
&= \frac{4\gamma}{D} + 2\kappa\left[\frac{f_{\rm A}}{N\ell_{\rm A}^2} + \frac{f_{\rm B}}{N\ell_{\rm B}^2}\right]D^2
\end{align*}

Now, to determine the rigidity at equilibrium, the equilibrium lamellar spacing is found by minimizing $\mathcal{F}^{\rm SST}(D)$ with respect to $D$:
\begin{align*}
\frac{{\rm d}\mathcal{F}^{\rm SST}}{{\rm d}D}\bigg|_{D^*} = 0 = -\frac{2\gamma}{(D^*)^2} + 2\kappa\left[\frac{f_{\rm A}}{N\ell_{\rm A}^2} + \frac{f_{\rm B}}{N\ell_{\rm B}^2}\right]D^*
\end{align*}
so 
\begin{align*}
D^* = \left(\frac{\gamma}{\kappa}\frac{N}{f_{\rm A}\ell_{\rm A}^{-2} + f_{\rm B}\ell_{\rm B}^{-2}}\right)^{1/3}
\end{align*}
The equilibrium rigidity is therefore
\begin{align*}
C_\parallel^{\rm SST}(D^*) &= \frac{4\gamma}{D^*} + 2\kappa\left[\frac{f_{\rm A}}{N\ell_{\rm A}^2} + \frac{f_{\rm B}}{N\ell_{\rm B}^2}\right](D^*)^2 \\ 
&= 4\gamma \left(\frac{\kappa}{\gamma}\frac{f_{\rm A}\ell_{\rm A}^{-2} + f_{\rm B}\ell_{\rm B}^{-2}}{N}\right)^{1/3} + 2\kappa\left[\frac{f_{\rm A}}{N\ell_{\rm A}^2} + \frac{f_{\rm B}}{N\ell_{\rm B}^2}\right]\left(\frac{\gamma}{\kappa}\frac{N}{f_{\rm A}\ell_{\rm A}^{-2} + f_{\rm B}\ell_{\rm B}^{-2}}\right)^{2/3} \\ 
&= 6 \left[\kappa\gamma^2\left(\frac{f_{\rm A}}{N\ell_{\rm A}^{2}} + \frac{f_{\rm B}}{N\ell_{\rm B}^{2}}\right)\right]^{1/3}
\end{align*}
Note that, following Helfand and Sapse~\cite{Helfand1975,Helfand1975_2} we can express $\gamma = k_BT\rho\sqrt{\ell_{\rm A}\ell_{\rm B}}\sqrt{\chi/6} = n_{\rm ch}k_B T N\sqrt{\ell_{\rm A}\ell_{\rm B}}\sqrt{\chi/6}$, allowing us to re-write the rigidity as
\begin{align*}
C_\parallel^{\rm SST}(D^*) &= \left(\frac{9\pi^2}{8}\right)^{1/3}\left(\chi N\right)^{1/3}n_{\rm ch}k_B T \left[\sqrt{N \ell_{\rm A}^2}\sqrt{N \ell_{\rm B}^2}\left(\frac{f_{\rm A}}{N\ell_{\rm A}^{2}} + \frac{f_{\rm B}}{N\ell_{\rm B}^{2}}\right)\right]^{1/3} \, .
\end{align*}


\begin{figure}[h!]
    \centering
    \includegraphics[width=0.9\linewidth]{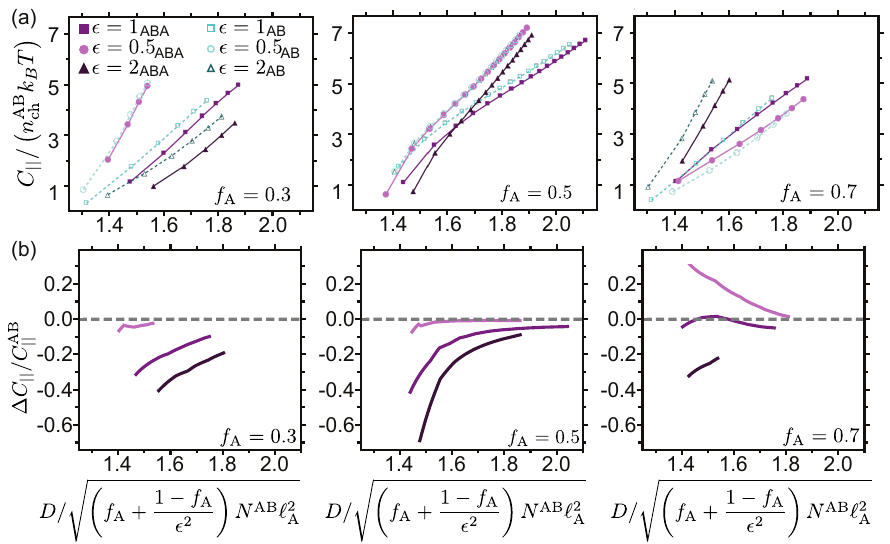}
    \caption{(a) Plot showing the scaling of longitudinal modulus with domain spacing for lamellae at $f_{\rm A}=0.3,0.5,\;\rm{and}\;0.7$, with domain spacing normalized by $\epsilon$ and $f_{\rm A}$. (b) Plot showing the percent difference between $C_{||}$ for ABA and AB, where $\Delta C_{||} = C_{||}^{\rm ABA}-C_{||}^{\rm AB}$.}
    \label{fig:lam_Dspacing_full}
\end{figure}

\section{Polycrystalline Averages}

Using the Voigt notation for symmetric tensors, we define linear elastic stiffness tensor components $C_{IJ} = C_{ijkl}$ associated with a given free energy $F$ and deformation $\strain$ as follows:
\begin{equation}
  \begin{split}
    C_{11}&=\frac{\partial^2F}{\partial \strain^2_{xx}}\quad \quad
    C_{22}=\frac{\partial^2F}{\partial \strain^2_{yy}}\quad \quad
    C_{33}=\frac{\partial^2F}{\partial \strain^2_{zz}}\quad \quad
    C_{44}=\frac{\partial^2F}{\partial \strain^2_{xy}}\quad \quad
    C_{55}=\frac{\partial^2F}{\partial \strain^2_{yz}}\quad \quad
    C_{66}=\frac{\partial^2F}{\partial \strain^2_{zx}} \\
    C_{12}&=C_{21}=\frac{\partial^2F}{\partial \strain_{xx}\partial \strain_{yy}}\quad \quad \quad \quad
    C_{23}=C_{32}=\frac{\partial^2F}{\partial \strain_{yy}\partial \strain_{zz}}\quad \quad \quad \quad \quad
    C_{13}=C_{31}=\frac{\partial^2F}{\partial \strain_{zz}\partial \strain_{xx}}
  \end{split}
\end{equation}

For the following, we note that the lamellar, columnar, and cubic phases can all be regarded as limiting cases of orthorhombic materials.
Following the method outlined by Berryman,~\cite{berryman_bounds_2011}, the Voigt and Reuss averages for overall shear response of orthorhombic systems is given by:
\begin{equation} \label{Gv}
    G_V=\frac{1}{5}\left[2G_{\rm eff}^v+C_{44}+C_{55}+C_{66}\right]
\end{equation}
and
\begin{equation} \label{Gr}
    \frac{1}{G_R}=\frac{1}{5}\left[\frac{2}{G_{\rm eff}^r}+\frac{1}{C_{44}}+\frac{1}{C_{55}}+\frac{1}{C_{66}}\right]\, ,
\end{equation}
respectively.
Here, the effective shear constants for orthorhombic systems can be defined as follows:
\begin{equation} \label{Gveff}
    G_{\rm eff}^v\equiv \frac{1}{6}\left[C_{11}+C_{22}+C_{33}-C_{12}-C_{13}-C_{23}\right]
\end{equation}
and
\begin{equation} \label{Greff}
\begin{split}
    \frac{3}{G_{\rm eff}^r}\equiv &\frac{2}{\Delta}\big[C_{11}(C_{22}+C_{33}) + C_{22}(C_{33}+C_{13}) + C_{33}(C_{11}+C_{12})\\
    &-C_{13}(C_{13}+C_{12})-C_{12}(C_{12}+C_{23})-C_{23}(C_{23}+C_{13})\big],
\end{split}
\end{equation}
where the factor $\Delta$ is given by
\begin{equation} \label{delta}
    \Delta= C_{11}C_{22}C_{33} +2C_{12}C_{23}C_{13} - C_{11}C_{23}^2 - C_{22}C_{13}^2 - C_{33}C_{12}^2,
\end{equation}
which is the pertinent determinant of the upper left $3\times3$ segment of the elastic constant matrix.

In order to apply these averages to our systems, we need to contend with two complications: (i) certain components $C_{IJ}$ are zero, which can result in singular evaluations of the above relations; (ii) since volumes are constrained to remain constant, the effective bulk modulus $K$ should be taken to be infinite.
To address this, we will \emph{regularize} both the vanishing moduli and diverging bulk moduli, taking the appropriate limits in the final evaluation. 


\subsection{1D polycrystals}
In the text, we note that under constant volume conditions (${\rm tr}\,\bstrain = 0$), the free energy for the lamellar phase can be written as
\begin{align*}
\Delta \mathcal{F} = \frac{1}{2}C_{||}\strain_{zz}^2\, .
\end{align*}
First, add the ability to change volume back in.
To respect the symmetries of the lamellae, we need add in a bulk modulus term $K$ as well as a transverse modulus $C_\perp$, as
\begin{align*}
\Delta \mathcal{F} = \frac{1}{2}\left[\left(C_{||} - C_\perp\right)\strain_{zz}^2 + C_\perp \left(\strain_{xx} + \strain_{yy}\right)^2 + K\left(\strain_{xx} + \strain_{yy} + \strain_{zz}\right)^2\right]\, ,
\end{align*}
where we note that we recover the original form in the limit that ${\rm tr}\,\bstrain \to 0$.
Note that the volume change will scale as ${\rm tr}\,\bstrain \sim 1/K$ under a finite pressure, so the ${\rm tr}\,\bstrain \to 0$ condition can be obtained in the $K \to \infty$ limit.

Let us re-write this as an orthorhombic system:
\begin{align*}
\Delta \mathcal{F} = \frac{1}{2}\big[C_{11}\strain_{xx}^2 + C_{22} \strain_{yy}^2 + C_{33} \strain_{yy}^2 + 2C_{12}\strain_{xx}\strain_{yy} + 2 C_{23}\strain_{yy}\strain_{zz} + 2 C_{31}\strain_{zz}\strain_{xx} + C_{44}\strain_{xy}^2 + C_{55}\strain_{yz}^2 + C_{66}\strain_{zx}^2 \big]
\end{align*}
Matching coefficients, we find
\begin{align*}
    C_{11} &= C_\perp + K \\
    C_{22} &= C_\perp + K \\
    C_{33} &= C_\parallel - C_\perp + K \\
    C_{12} &= C_\perp + K + \varepsilon \\
    C_{23} &= K \\
    C_{31} &= K \\
    C_{44} &= C_{55} = C_{66} = \delta 
\end{align*}
where we have introduced constants $\varepsilon$ and $\delta$ as regularizing constants, such that the correct coefficient relations are found in the limit $\varepsilon,\delta \to 0$.
Now, we evaluate the terms needed to compute the polycrystalline averages: 
\begin{align*}
G_{\rm eff}^v &= \frac{1}{6}\left(C_{\parallel} - \varepsilon \right)\\ 
G_V &= \frac{1}{15}\left(C_\parallel + 9\delta - \varepsilon\right) \\
\Delta &= 2(C_\perp^2 - KC_\parallel - C_\parallel C_\perp)\varepsilon + \mathcal{O}(\varepsilon^2) \\
G_{\rm eff}^r &= \frac{-3(C_\perp^2 - KC_\parallel - C_\parallel C_\perp^2)}{KC_\perp - 4 K C_\parallel + 4 C_\perp^2 - 4 C_\perp C_\parallel}\varepsilon + \mathcal{O}(\varepsilon^2) \\
G_R &= \frac{-15(C_\perp^2 - KC_\parallel - C_\parallel C_\perp^2)}{2 K C_\perp - 8( K C_\parallel - C_\perp^2 + C_\perp C_\parallel)}\varepsilon + \mathcal{O}(\varepsilon^2)
\end{align*}
Note that the $K\to \infty$ limit gives finite results for $G_V$ and $G_R$.
Finally, taking the $\varepsilon,\delta \to 0$ limits, we find
\begin{equation}
\begin{split}
    G_V &= \frac{1}{15}C_\parallel \\
    G_R &= 0
\end{split}
\end{equation}

\subsection{2D polycrystals}
The energy for 2D hexagonal cylinders is given by
\begin{equation}\label{2d_polycrystal}
    \Delta  \mathcal{F} = \frac{1}{2}\left[2\mu_{\perp} (\strain_{xx}^2 + \strain_{yy}^2 + 2\strain_{xy}\strain_{yx}) + \lambda_{\perp}(\strain_{xx} + \strain_{yy})^2\right]
\end{equation}
Once again, we modify this form by adding in terms that should disappear as ${\rm tr}\,\bstrain \to 0$:
\begin{align*}
    \Delta\mathcal{F} = \frac{1}{2}\left[2\mu_{\perp} (\strain_{xx}^2 + \strain_{yy}^2 + 2\strain_{xy}\strain_{yx}) + \eta \strain_{zz}^2 + \left(\lambda_{\perp} - \eta\right)(\strain_{xx} + \strain_{yy})^2 + K\left(\strain_{xx} + \strain_{yy} + \strain_{zz}\right)^2\right]
\end{align*}
where $K$ is the bulk modulus and $\eta$ describes the stiffness along the $z$-axis.
Matching coefficients, we find
\begin{align*}
    C_{11} &= 2\mu_\perp + \lambda_\perp - \eta + K  \\
    C_{22} &= 2\mu_\perp + \lambda_\perp - \eta + K \\
    C_{33} &= \eta + K \\
    C_{12} &= 2\mu_\perp + \lambda_\perp - \eta + K + \varepsilon \\
    C_{23} &= K \\
    C_{31} &= K \\
    C_{44} &= 4\mu_\perp \\
    C_{55} &= C_{66} = \delta 
\end{align*}
Now, we evaluate the terms needed to compute the polycrystalline averages: 
\begin{align*}
G_{\rm eff}^v &= \frac{1}{6}\left(2\mu_\perp + \lambda_\perp - \varepsilon \right)\\ 
G_V &= \frac{1}{15}\left(14\mu_\perp + \lambda_\perp - \varepsilon + 6\delta \right) \\
\Delta &= 2\left((\eta - 2\mu_\perp-\lambda_\perp)\eta - (2\mu_\perp + \lambda_\perp)K\right)\varepsilon + \mathcal{O}(\varepsilon^2) \\
G_{\rm eff}^r &= -\frac{3((2\mu_\perp + \lambda_\perp)(\eta + K) - \eta^2)}{4(2\mu_\perp + \lambda_\perp - \eta)\eta + (3(2\mu_\perp + \lambda_\perp) + \eta)K}\varepsilon + \mathcal{O}(\varepsilon^2) \\
G_R &= -\frac{15((2\mu_\perp + \lambda_\perp)(\eta + K) - \eta^2)}{8(2\mu_\perp + \lambda_\perp - \eta)\eta + 2(3(2\mu_\perp + \lambda_\perp) + \eta)K}\varepsilon + \mathcal{O}(\varepsilon^2)
\end{align*}
Note that the $K\to \infty$ limit gives finite results for $G_V$ and $G_R$.
Finally, taking the $\varepsilon,\delta \to 0$ limits, we find
\begin{equation}
\begin{split}
    G_V &=\frac{1}{15}\left(14\mu_\perp + \lambda_\perp\right) = \frac{1}{15}\left(13\mu_\perp + Y_\parallel\right) \\
    G_R &=0
\end{split}
\end{equation}


\subsection{3D polycrystals}
The free energy per volume for cubic systems is given by
\begin{equation}
\begin{split}
    \Delta \mathcal{F} &= \frac{Y}{9}\left((\strain_{xx} - \strain_{yy})^2 + (\strain_{yy} - \strain_{zz})^2 + (\strain_{zz} - \strain_{xx})^2\right) \\
    &\mkern+16mu + 2\mu\left(\strain_{xy}\strain_{yx} + \strain_{yz}\strain_{zy} + \strain_{zx}\strain_{xz}\right) + \frac{1}{2}K(\strain_{xx} + \strain_{yy} + \strain_{zz})^2\, ,
\end{split}
\end{equation}
which we extend to include a bulk rigidity.
Matching coefficients, we find
\begin{align*}
    C_{11} &= C_{22} = C_{33} = \frac{4}{9}Y + K  \\
    C_{12} &= C_{23} = C_{31} = -\frac{2}{9}Y + K \\
    C_{44} &= C_{55} = C_{66} = 4\mu 
\end{align*}
Now, we evaluate the terms needed to compute the polycrystalline averages: 
\begin{align*}
G_{\rm eff}^v &= \frac{1}{3}Y\\ 
G_V &= \frac{2}{15}\left(18\mu + Y\right)\\
\Delta &= \frac{4}{3}Y^2K \\
G_{\rm eff}^r &= \frac{27 Y K}{4Y + 90 K} \\
G_R &= \frac{540 \mu Y K}{32\mu Y + 9(80 \mu + 9 Y)K}
\end{align*}
Taking the $K \to \infty$ limit, we find
\begin{equation}
\begin{split}
    G_V &= \frac{2}{15}\left(18\mu + Y\right)  \\
    G_R &= \frac{60 \mu Y}{80\mu + 9 Y}
\end{split}
\end{equation}


\begin{figure}[h!]
    \centering
    \includegraphics[width=0.8\linewidth]{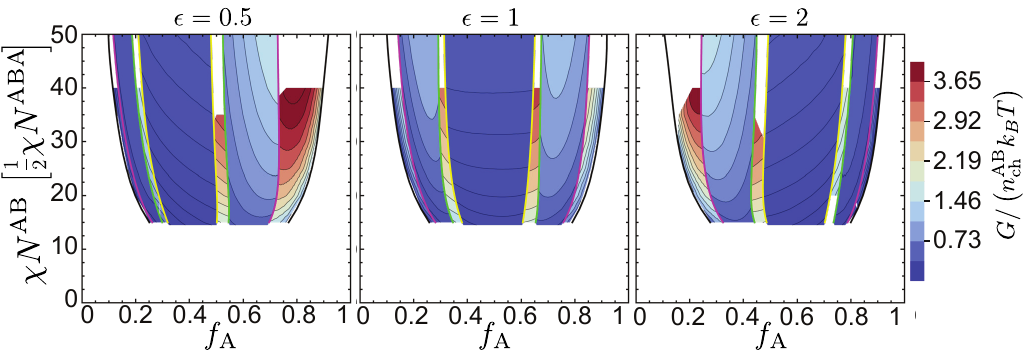}
    \caption{Heat map of Voigt (upper) modulus bounds for each $\epsilon$ for ABA.}
    \label{fig:composite_heat_map}
\end{figure}

\begin{figure}[h!]
    \centering
    \includegraphics[width=0.8\linewidth]{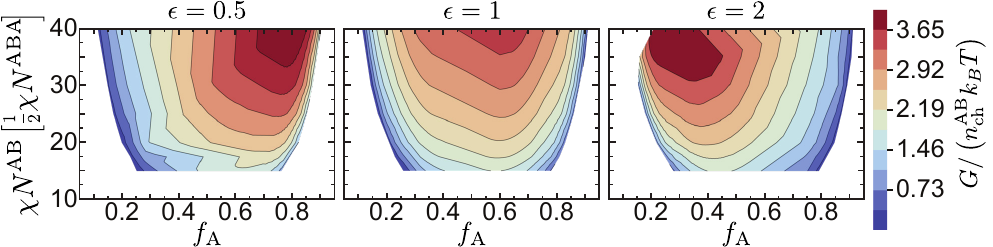}
    \caption{Convex hull for each $\epsilon$ for ABA.}
    \label{fig:convex_hull}
\end{figure}

\eject
\section{Comparisons of moduli with literature values}
\begin{table}[h!]
    \centering
    \small
    \begin{tabular}{|cccccc|ccc|}
    \hline
        \multirow{2}{4em}{Polymer} & \multirow{2}{1em}{$f_{\rm A}$} & \multirow{2}{3em}{Phase} & \multirow{2}{1em}{$N$} & \multirow{2}{1.5em}{$\chi N^a$} & \multirow{2}{3em}{$T$ ($^\circ$C)} & Exp. & Theor. & Theor. \\
        &&&&&& $G'$ (kPa) & $G'_{\epsilon=1}$  & $G'_{\epsilon \neq 1}$ \\
        \hline
        PI-PDMS$^b$ & 0.47 & Lam & 257 & 12.9 / 12.5 & -12 & 0.35 & 0--14 & 0--13$^d$ \\
        PI-PDMS & 0.64 & Gyr & 490 & 24.5 / 30 & 120 & 40 & 47--263 & 19--140$^d$ \\
        PEP-PDMS$^c$ & 0.49 & Lam & 384 & 5.0 / 12.5 & 62 & 0.08 & 0--12 & 0--11$^d$ \\
        PEP-PDMS & 0.66 & Gyr & 800 & 10.4 / 30 & 160 & 30 & 25--171 & 6--118$^d$ \\
        \hline
        PS-PI$^e$ & 0.17 & Hex & 830 & 35 / 30 & 135 & 1.1 & 0--18 & 0--12$^f$ \\
        PS-PI & 0.17 & BCC & 830 & 30 / 30 & 180 & 6 & 5--74 & 3--47$^f$ \\
        PS-PI-PS & 0.17 & Hex & 1660 & 60$^g$ / 30 & 180 & 2.5 & 0--21 & 0--13$^f$ \\
        PS-PI-PS & 0.17 & BCC & 1660 & 57$^g$ / 30 & 210 & 4 & 3.4--77 & 1.7--44$^f$ \\
        \hline
        \multicolumn{9}{l}{$^a$experimental / theoretical $\chi N$ used. $^b\epsilon=1.3$. $^c\epsilon=1.5$. $^d\epsilon=2$, values from Almdal et al.~\cite{almdal1996order}} \\
        \multicolumn{9}{l}{$^e\epsilon=0.67$. $^f\epsilon=0.5$, values from Ryu et al~\cite{ryu1997structure} $^g$not normalized to AB chains}\\
    \end{tabular}
    \caption{Comparison between experimental and polycrystalline average moduli, with relevant design parameters for each polymer.}
    \label{tab:litcomparison}
\end{table}

\eject
\section{Polymer bridging calculations}

\begin{figure}[h!]
    \centering
    \includegraphics[width=1.0\linewidth]{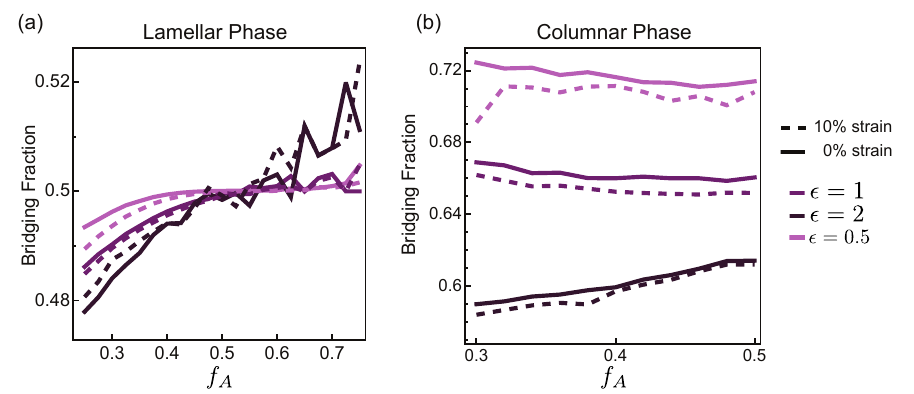}
    \caption{Bridging fractions for ABA triblocks for lamellar and columnar phases.}
    \label{fig:bridging_fraction}
\end{figure}

In ABA triblock copolymers, we quantify the fraction of bridging chains following a standard chain propagator framework~\cite{Matsen1999}. 
For each parameter set ($f_{\rm A}$, $\epsilon$, $\chi$N), we first obtain the self-consistent chemical potential fields $\omega_{\rm A}(r)$, $\omega_{\rm B}(r)$ using SCFT. We then compute the forward and backward single chain propagators $q(r,s)$ and $q^\dagger(r,s)$ by integrating the modified diffusion equation along the contour length s $\in [0,1]$,
\begin{align}
\partial_s q(r,s) = \frac{l^2_{j(s)}}{6}\nabla^2q(r,s) \;-\; w_{j(s)}(r)q(r,s),
\label{mod. diff. eqn}
\\
-\partial_s q^\dagger(r,s) = \frac{l^2_{j(s)}}{6}\nabla^2q^\dagger(r,s) \;-\; w_{j(s)}(r)q^\dagger(r,s)
\end{align}
with initial/terminal conditions $q(r,0) = 1$ and $q^\dagger(r,1) = 1$. The block label $j(s) \in \{A,B\}$ is defined as,
\[
j(s) = \begin{cases}
B, & f_1 \le s \le f2\\
A, & \text{otherwise}
\end{cases}
\]
Here $f_1$ and $f_2$ are the contour length coordinates of the B-block entry and exit respectively, with $0<f_1<f_2<1$.
With Q as the single chain partition function, the segment probability density is defined as,
\begin{align}
\rho(r,s) = \frac{q(r,s)q^\dagger(r,s)}{Q}
\end{align}
Now, to further resolve these chains as looping or bridging, we focus on the B block, i.e. $s\in[f1,f2]$, that spans the midblock. Starting from the B-block entrance (s=f1), we construct a function $\bar q(r,s)$ as follows,
\begin{align}
    \bar q(r,s) = q(r,f_1); r\in 1^{\text{st}} \text{ unit cell},
\end{align}
and 0 elsewhere. We propagate this function using the same modified diffusion equation \eqref{mod. diff. eqn} from $s = f_1$ to $s = f_2$, but with Neumann boundary conditions at the ends of the domain. We can therefore calculate the distribution $\bar \rho(r,s)$ at the end of the B block domain by
\begin{align}
    \bar \rho(r,f_2) = \frac{\bar q(r,f_2)q^\dagger(r,f_2)}{Q}
\end{align}
The looping fraction $\nu_L$ can be calculated by integrating $\bar \rho(r,f_2)$ over the unit cell, i.e. chains that left the A domain, traversed the B block, and returned back to the same A domain:
\begin{align}
    \nu_L = \int_{\Omega_{\text{cell}}}\bar \rho(r,f_2)\,dr,\\
    \text{and } \nu_B = 1 - \nu_L
\end{align}
Here, $\nu_B$ is the bridging fraction. The bridging fractions for lamellar and columnar phases for a range of A-block fractions ($f_\text{A}$) are reported in Figure~\ref{fig:bridging_fraction}. \\

\remove{To further probe how the chain segments rearrange with strain, we evaluate the spatial profile of a single chain using these propagators. For a chosen contour position $s^*$, we normalize the segment probability density over one unit cell (domain size D) to obtain the distribution of the chain segment $s^*$,}
\begin{align}
    g(x/D) \;=\; \frac{\rho(x,s^*)}{\displaystyle \int_0^{D}\rho(x, s^*)\,dx},
\end{align}
\remove{so that $\int_0^1 g(x/D)\,dx = 1$. For AB diblocks we evaluate the chain end segment in the B domain ($s^* = 1$), whereas for ABA triblocks we pick the mid-chain segment ($s^*=0.5$) of the B-block. 
For each $\epsilon$, we plot the undeformed (0\% strain) and deformed (10\% strain) profiles, and compare the AB and ABA on the same axis (see Figure~\ref{fig:g(x/D)}).}

\edits{\section{Domain spacing}}

\begin{figure}[h!]
    \centering
    \includegraphics[width=0.5\linewidth]{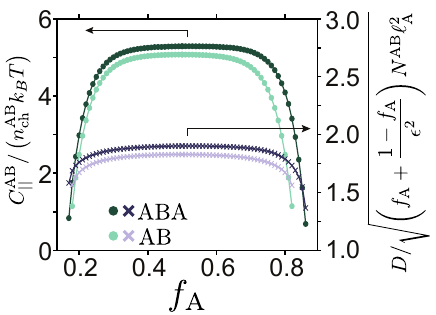}
    \caption{\edits{Comparison of longitudinal modulus $C_{||}$ and domain spacing $D$ with respect to $f_{\rm A}$, comparing ABA and AB.}}
    \label{fig:placeholder}
\end{figure}

\section*{References}
\bibliography{refs}